\pdfoutput=1
\documentclass[runningheads]{llncs}
\usepackage{etex} 

\usepackage{url}
\usepackage{amsmath,amsfonts,amssymb}
\usepackage{nccmath} 
\usepackage{fancyvrb}
\usepackage{stmaryrd}
\usepackage[inline]{enumitem}
\usepackage{etoolbox}
\usepackage[ruled,vlined,linesnumbered]{algorithm2e}
\SetKwRepeat{Do}{do}{while}
\SetKwComment{tcp}{$\rhd\;$}{}%
\makeatletter
\patchcmd{\@algocf@start}
{-1.5em}
{0pt}
{}{}
\makeatother

\usepackage{pifont}
\newcommand{\cmark}{\textcolor{ForestGreen}{\ding{51}}}%
\newcommand{\xmark}{\textcolor{Maroon}{\ding{55}}}%
\newcommand{\namark}{\textcolor{Maroon}{\textbf{?}}}%

\usepackage[font=small,labelfont=bf]{caption}
\usepackage{comment}
\usepackage[long]{optidef}
\usepackage{algpseudocode}
\usepackage{float}
\usepackage{graphicx}
\usepackage{adjustbox}
\usepackage{cite}
\usepackage{booktabs}
\usepackage[usenames,dvipsnames]{xcolor}

\usepackage{nicematrix}

\NiceMatrixOptions{cell-space-top-limit = 1pt, cell-space-bottom-limit = 1pt}

\usepackage{array}
\usepackage{multirow}
\usepackage{bm}
\usepackage{marvosym}
\usepackage{fontawesome}

\usepackage{graphicx}
\usepackage{wrapfig}
\usepackage{subfig}

\usepackage[title]{appendix}

\usepackage{scalerel} 

\patchcmd{\paragraph}{\itshape}{\bfseries\boldmath}{}{} 

\patchcmd\subequations
{\theparentequation\alph{equation}}
{\subequationsformat}
{}{}

\makeatletter
\def\thanks#1{\protected@xdef\@thanks{\@thanks
		\protect\footnotetext{#1}}}
\makeatother

\makeatletter
\RequirePackage[bookmarks,unicode,colorlinks=true]{hyperref}%
\def\@citecolor{blue}%
\def\@urlcolor{blue}%
\def\@linkcolor{RedViolet}%

\def\orcidID#1{\smash{\href{http://orcid.org/#1}{\protect\raisebox{-1.25pt}{\protect\includegraphics{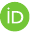}}}}}
\makeatother

\newcommand{\subequationsformat}{\theparentequation.\arabic{equation}}

\include{def}

\newif\ifcameraready
\camerareadyfalse

\begin{document}
\title{Synthesizing Invariant Barrier Certificates via Difference-of-Convex Programming\thanks{\setlength{\leftskip}{0em}%
	This work has been partially funded by the NSFC under grant No.~61625206, 61732001, 61872341, and 61836005, by the ERC Advanced Project FRAPPANT under grant No.~787914, and by the CAS Pioneer Hundred Talents Program.
	}
}
\titlerunning{Synthesizing Invariant Barrier Certificates via DCP}
%


\authorrunning{Q.~Wang et al.}

\author{Qiuye Wang\inst{1,2}$^{\text{(\Letter)}}$\orcidID{0000-0001-5138-3273}
	\and
	Mingshuai Chen\inst{3}$^{\text{(\Letter)}}$\orcidID{0000-0001-9663-7441}
	\and
	Bai Xue\inst{1,2}$^{\text{(\Letter)}}$\orcidID{0000-0001-9717-846X} 
	\and
	\\Naijun Zhan\inst{1,2}$^{\text{(\Letter)}}$\orcidID{0000-0003-3298-3817}
	\and
	Joost-Pieter~Katoen\inst{3}$^{\text{(\Letter)}}$\orcidID{0000-0002-6143-1926}
}

\institute{
	SKLCS, Institute of Software, CAS, Beijing, China
	\and
	University of Chinese Academy of Sciences, Beijing, China\\
	\email{\{wangqye,xuebai,znj\}@ios.ac.cn}
	\and
	RWTH Aachen University, Aachen, Germany\\
	\email{\{chenms,katoen\}@cs.rwth-aachen.de}
}

\maketitle

\setcounter{footnote}{0}


\setlength{\floatsep}{1\baselineskip}
\setlength{\textfloatsep}{1\baselineskip}
\setlength{\intextsep}{1\baselineskip}

\begin{abstract}
    A barrier certificate often serves as an inductive invariant 
    that isolates an unsafe region from the reachable set of states, 
    and hence is widely used in proving safety of hybrid systems
    possibly over the infinite time horizon. 
    We present a novel condition on barrier certificates, 
    termed the \emph{invariant barrier-certificate condition}, 
    that witnesses unbounded-time safety of differential dynamical systems. 
    The proposed condition is by far the least conservative one on barrier certificates,  and can be shown as the weakest possible one to attain inductive invariance. 
    We show that discharging the invariant barrier-certificate condition 
    ---thereby synthesizing invariant barrier certificates--- 
    can be encoded as solving an \emph{optimization problem subject to bilinear matrix inequalities} (BMIs). 
    We further propose a synthesis algorithm based on difference-of-convex programming, 
    which approaches a local optimum of the BMI problem 
    via solving \emph{a series of convex optimization problems}. 
    This algorithm is incorporated in a branch-and-bound framework 
    that searches for the global optimum in a divide-and-conquer fashion. 
    We present a weak completeness result of our method, 
    in the sense that a barrier certificate is guaranteed to be found (under some mild assumptions) 
    whenever there exists an inductive invariant (in the form of a given template) 
    that suffices to certify safety of the system. 
    Experimental results on benchmark examples demonstrate the effectiveness and efficiency of our approach. 
\end{abstract}

\section{Introduction}\label{sec:introduction}

Hybrid systems are mathematical models that capture the interaction 
between continuous physical dynamics and discrete switching behaviors, 
and hence are widely used in modelling cyber-physical systems (CPS). 
These CPS may be complex and safety-critical, 
with sensitive variables of the environment in its sphere of control. 
Everyday examples include process control at all scales, 
ranging from household appliances to nuclear power plants, 
or embedded systems in transportation domain, 
such as autonomous driving maneuvers in automotive, 
aircraft collision-avoidance protocols in avionics, 
or automatic train control applications, 
as well as a broad range of devices in health technologies, such as cardiac pacemakers.

The safety-critical feature of these CPS, with increasingly complex behaviors, has initiated automatic safety 
or, dually, reachability verification of hybrid systems~\cite{ACH95,DBLP:journals/siglog/Fraenzle19}. 
The problem of reachability verification is undecidable in general~\cite{ACH95}, 
albeit with decidable families of sub-classes 
(see, e.g.,~\cite{LPY01,DBLP:conf/hybrid/AnaiW01,Gan15,DBLP:conf/eucc/GanCLXZ16,Gan18}) 
identified in the literature. 
The hard core of the verification problem lies in 
reasoning about the continuous dynamics, 
which are often characterized by ordinary differential equations (ODEs). 
In particular, when nonlinearity arises in the ODEs, 
the explicit computation of the exact reachable set is usually intractable 
even for purely continuous dynamics~\cite{WDSmith06}.

Therefore in the literature, a plethora of approximation schemes, 
as surveyed in~\cite{DBLP:journals/siglog/Fraenzle19}, 
for reachability analysis of hybrid systems has been developed, 
including an invariant-style reasoning scheme known as \emph{barrier certificate}~\cite{Prajna04}. 
A barrier certificate often serves as an inductive invariant that isolates an unsafe region from the reachable set, 
thereby witnessing safety of hybrid systems possibly over the infinite time horizon. 
A common way to synthesize barrier certificates is 
to reduce the condition defining barrier certificates 
to a numerical optimization or constraint solving problem. 
There is, however, a trade-off between the expressiveness of the barrier-certificate condition 
and the efficiency in discharging the reduced constraints. 
Hence, to enable efficient algorithmic synthesis of barrier certificates 
via, e.g., linear programming (LP), second-order cone programming (SOCP), semidefinite programming (SDP) 
and interval analysis~\cite{djaballah2017construction,YANG2020100837,DBLP:conf/cav/KongSG18}, 
the general condition on inductive invariance 
(that a barrier certificate defines an invariant, see~\cite{Platzer18FM,Gan17}) 
has been strengthened into a spectrum of different shapes, 
e.g., \cite{Kong13, yang2015exact, zeng2016darboux, Gan17, Platzer18FM}. 
It has been, nevertheless, a long-standing challenge 
\emph{to find a barrier-certificate condition that is as weak as possible 
while admitting efficient synthesis algorithms}.

In this paper, we present a new condition on barrier certificates, 
termed the \emph{invariant barrier-certificate condition}, 
based on the sufficient and necessary condition on being an inductive invariant~\cite{LZZ11}. 
Our invariant barrier-certificate condition is by far, to the best of our knowledge, the least conservative one on barrier certificates, 
and can be shown as the weakest possible one to attain inductive invariance. 
We show, by leveraging Putinar's Positivstellensatz~\cite{lasserre2010moments}, 
that discharging the invariant barrier-certificate condition 
---thereby synthesizing invariant barrier certificates--- 
can be encoded as solving an optimization problem subject to \emph{bilinear matrix inequalities} (BMIs). 
%
%
We further show that general bilinear matrix-valued functions 
can be decomposed as a difference 
of two psd-convex (extension of convexity to matrix-valued functions) functions using eigendecomposition, 
thus resulting in a synthesis algorithm 
as per \emph{difference-of-convex programming} (DCP)~\cite{tao1986algorithms,le2018dc}, 
which solves a series of convex sub-problems (in the form of \emph{linear matrix inequalities} (LMIs)) 
that approaches (arbitrarily close to) a local optimum of the BMI problem. 
This algorithm is incorporated in a branch-and-bound framework
that searches for the global optimum in a divide-and-conquer fashion. 
We present a weak completeness result of our method, 
in the sense that a barrier certificate is guaranteed to be found (under some mild assumptions) 
whenever there exists an inductive invariant (in the form of a given template) 
that suffices to certify the system's safety. 
A similar result on completeness is previously provided only by symbolic approaches, 
yet to the best of our knowledge, 
not by methods base on numerical constraint solving, e.g.,~\cite{yang2015exact, yang2016linear, CAV20BMI}. 
Experiments on a collection of examples suggested that 
our invariant barrier-certificate condition recognizes more barrier certificates than existing conditions, 
and that our DCP-based algorithm is more efficient than directly solving the BMIs via off-the-shelf solvers.
\ifcameraready
\else
We defer all proofs to Appendix~\ref{appendix_proofs}.
\fi

\section{A Bird's-Eye Perspective}\label{sec:overview}

We use the following example to give a bird's-eye view of our approach.

\begin{example}[\expname{overview}~\textnormal{\cite{djaballah2017construction}}]\label{exmp:overview}
Consider the following continuous-time dynamical system modelled by an ordinary differential equation:
\begin{equation*}
	\dot{\xx} =
    \begin{pmatrix}
        \dot{x}_1 \\
        \dot{x}_2
    \end{pmatrix} 
    =
    \begin{pmatrix}
        x_1 + x_2 \\
        x_1 x_2 - 0.5 x_2^2 + 0.1 
    \end{pmatrix}
	.
\end{equation*}%
The verification obligation is to show that 
the system trajectory originating from any state in the initial set 
$\init = \{ \xx \mid \initBound(\xx) \leq 0 \}$ with $\initBound(\xx) = x_1^2 + (x_2 - 2)^2 -1$ 
will never enter the unsafe set $\unsafe = \{ \xx \mid \unsafeBound(\xx)\leq 0 \}$ 
with $\unsafeBound(\xx) = x_2 + 1$.
\qedTT
\end{example}

A barrier certificate satisfying our condition in Definition~\ref{def:invBc} 
serves as an inductive invariant that suffices to isolate the unsafe region $\unsafe$ 
from the set of reachable states from $\init$, 
thereby proving safety of the system over the infinite time horizon. 
To this end, we proceed in the following steps.

\paragraph*{1) Encode as Sum-of-Squares (SOS) Constraints.}
We set a (polynomial) barrier-certificate template $B(\aaa, \xx) = a x_2$ 
with unknown coefficient $a \in \mathbb{R}$. 
According to Theorem~\ref{thm:invariantCondition}, 
we only need to consider Lie derivatives up to order $\LieBound = 1$, 
i.e., $\mathcal{L}_{\ff}^0 B(\aaa, \xx) = a x_2$ 
and $\mathcal{L}_{\ff}^1 B(\aaa, \xx) = a (x_1 x_2 - 0.5 x_2^2 + 0.1)$.

By Theorem~\ref{thm:invariantBcSosSufficient}, 
$B(\aaa, \xx)$ is an invariant barrier certificate 
if there exists a polynomial $v(\xx)$, SOS polynomials $\sigma(\xx), \sigma'(\xx)$ 
and a constant $ \epsilon > 0$ such that
\begin{subequations}\label{eqn:expInvCond}
\begin{align}
    &-\underbrace{a x_2}_{B} +\, \sigma(\xx) \underbrace{\left(x_1^2 + (x_2 - 2)^2 - 1\right)}_{\initBound}, \owntag[expInvCond1]{initial} \\[.1cm]
    &-\underbrace{a \left(x_1 x_2 - 0.5 x_2^2 + 0.1\right)}_{\mathcal{L}_{\ff}^1 B} + \,v(\xx) \underbrace{a x_2}_{\mathcal{L}_{\ff}^0 B}, \owntag[expInvCond2]{Lie consecution} \\[.1cm]
    & \underbrace{a x_2}_{B} +\, \sigma'(\xx) \underbrace{(x_2 + 1)}_{\unsafeBound} - \epsilon \owntag[expInvCond3]{separation}
\end{align}
\end{subequations}%
are SOS polynomials. We set $\epsilon = 0.01$ in this example. 

\paragraph*{2) Reduce to a BMI Optimization Problem.}
Observe that the above SOS constraints can be formulated as BMI constraints. 
For instance, let us assume that \eqref{eqn:expInvCond2} is an SOS polynomial of degree at most 2 
and $v(\sss, \xx) = s_0 + s_1 x_1 + s_2 x_2$ is a template polynomial with unknown coefficients $\sss$. 
Then constraint \eqref{eqn:expInvCond2} is equivalent to the BMI constraint
\begin{equation*}
    \mathcal{F}_2(\aaa, \sss) = -
    \begingroup 
    \setlength\arraycolsep{3pt}
	    \begin{pmatrix}
	        -0.1 a & 0 & 0.5a s_0 \\
	        0 & 0 & 0.5(a s_1 - a) \\
	        0.5a s_0 & 0.5(a s_1 - a) & a s_2 + 0.5a 
	    \end{pmatrix}
	\endgroup
    \preceq 0
\end{equation*}%
meaning that the bilinear matrix (LHS of $\preceq$) is negative semidefinite. 
Note that the bilinearity arises due to the coupling of the unknown coefficients $\aaa$ and $\sss$.

Constraints \eqref{eqn:expInvCond1} and \eqref{eqn:expInvCond3} can be reduced 
to BMI constraints in an analogous way\footnote{
Despite that no bilinearity is involved in constraints \eqref{eqn:expInvCond1} and \eqref{eqn:expInvCond3}, 
they can be processed in the same way as \eqref{eqn:expInvCond2}, yielding LMI constraints.}, 
yielding $\mathcal{F}_1$ and $\mathcal{F}_3$. 
It then follows that, to solve the SOS constraints, 
we need to find a feasible solution $(\aaa, \sss)$ such that\footnote{
Extra constraints on $\sigma(\xx)$ and $\sigma'(\xx)$ being SOS polynomials 
can be encoded analogously in the feasibility problem, yet are omitted here for the sake of simplicity.}
\begin{equation}\label{eqn:bmiExpFeasible}
    \mathcal{F}_1(\aaa, \sss) \preceq 0 
    \land \mathcal{F}_2(\aaa, \sss) \preceq 0 
    \land \mathcal{F}_3(\aaa, \sss) \preceq 0.
\end{equation}%

To exploit well-developed optimization techniques, 
the feasibility problem \eqref{eqn:bmiExpFeasible} is transformed 
to an optimization problem subject to BMI constraints:
\begin{maxi}
    {\lambda, \aaa, \sss}
    {\lambda}
    {\label{eqn:bmiExp}}
    {}
    \addConstraint{\mathcal{B}_i(\lambda, \aaa, \sss) \define \mathcal{F}_i(\aaa, \sss) + \lambda I}{\preceq 0,\quad}{i=1, 2, 3}
\end{maxi}%
where $I$ is the identity matrix with compatible dimensions. 
Note that problem \eqref{eqn:bmiExpFeasible} has a feasible solution 
if and only if the optimal value $\lambda^*$ in \eqref{eqn:bmiExp} is non-negative.

\paragraph*{3) Decompose as Difference-of-Convex Problems.}
The problem \eqref{eqn:bmiExp} contains non-convex constraints 
and hence does not admit efficient (polynomial-time) algorithms tailored for convex optimizations. 
However, by our technique presented in Section~\ref{sec:algorithm}, 
a non-convex function $\mathcal{B}_i(\lambda, \aaa, \sss)$ can be decomposed 
as the difference of two psd-convex (defined later) matrix-valued functions: 
\begin{equation}\label{eq:diff-of-psd-convex}
    \mathcal{B}_i(\lambda, \aaa, \sss) = 
    \mathcal{B}_i^+(\lambda, \aaa, \sss) - \mathcal{B}_i^-(\lambda, \aaa, \sss).
\end{equation}%
The decomposition of $\mathcal{B}_2(\lambda, \aaa, \sss)$, for instance, gives
\begin{align*}
    &\medmath{\mathcal{B}_2^+(\lambda, \aaa, \sss) =}  \\
    &\qquad\textstyle{\frac{1}{8}}\begin{pNiceMatrix}[small]
        8 \lambda + 0.08 a + a^2 + 0.408 s_0^2 & 
        0.408 s_0 s_1 & 
        -2 a s_0 + 0.816 s_0 s_2 \\
        0.408 s_0 s_1 & 
        8 \lambda + a^2 + 0.408 s_1^2 & 
        4 a - 2 a s_1 + 0.816 s_1 s_2 \\
        -2 a s_0 + 0.816 s_0 s_2 & 
        4 a - 2 a s_1 + 0.816 s_1 s_2 & 
        8 \lambda - 4 a + 2.449 a^2 - 4 a s_2 + s_0^2 + s_1^2 + 1.632 s_2^2 
    \end{pNiceMatrix} \\
	&\medmath{\mathcal{B}_2^-(\lambda, \aaa, \sss) =} \\
    &\qquad\textstyle{\frac{1}{8}}\begin{pNiceMatrix}[small]
        a^2 + 0.408 s_0^2 & 
        0.408 s_0 s_1 & 
        2 a s_0 + 0.816 s_0 s_2 \\
        0.408 s_0 s_1 & 
        a^2 + 0.408 s_1^2 & 
        2 a s_1 + 0.816 s_1 s_2 \\
        2 a s_0 + 0.816 s_0 s_2 & 
        2 a s_1 + 0.816 s_1 s_2 & 
        2.449 a^2 + 4 a s_2 + s_0^2 + s_1^2 + 1.632 s_2^2 
    \end{pNiceMatrix} 
	.
\end{align*}%
\vspace*{.01mm}

\paragraph*{4) Solve a Series of Convex Sub-problems.}
Now, we apply a standard iterative procedure 
in difference-of-convex programming~\cite{dinh2011combining} as follows. 
Given a feasible solution $\zz^k = (\lambda^k, \aaa^k, \sss^k)$ 
to the BMI optimization problem \eqref{eqn:bmiExp}, 
the concave part $-\mathcal{B}_i^-(\lambda, \aaa, \sss)$ in \eqref{eq:diff-of-psd-convex} 
is linearized around $\zz^k$, thus yielding a series of convex programs ($k = 0, 1, \ldots$): 
\begin{maxi}
    {\lambda, \aaa, \sss}
    {\lambda}
    {\label{eqn:bmiExpLinearized}}
    {}
    \addConstraint{\mathcal{B}_i^+(\zz) - \mathcal{B}_i^-\left(\zz^k\right) - \mathcal{DB}_i^-\left(\zz^k\right)\left(\zz -\zz^k\right)}{\preceq 0,\quad}{i=1, 2, 3}
\end{maxi}%
where $\mathcal{DB}_i^-$ denotes the derivative of the matrix-valued function $\mathcal{B}_i^-$. 

The soundness of our approach 
asserts that the feasible set of the linearized program \eqref{eqn:bmiExpLinearized} 
under-approximates the feasible set of the original BMI program \eqref{eqn:bmiExp}. 
Therefore, if $\lambda^k \geq 0$ after iteration $k$, 
we can safely claim that $(\aaa^k, \sss^k)$ is a feasible solution to \eqref{eqn:bmiExpFeasible}. 
A barrier certificate $B(\xx)$ is then obtained by substituting $\aaa^k$ in $B(\aaa, \xx)$. 
Moreover, if we take the optimum $\zz^{*, k}$ of \eqref{eqn:bmiExpLinearized} 
to be the next linearization point $\zz^{k+1}$, 
the solution sequence $\{ \zz^k \}_{k \in \NN}$ converges to a local optimum of \eqref{eqn:bmiExp}.

\begin{wrapfigure}{r}{0.47\textwidth}
	\vspace*{-8mm}
	\begin{center}
		\includegraphics[width=0.47\textwidth]{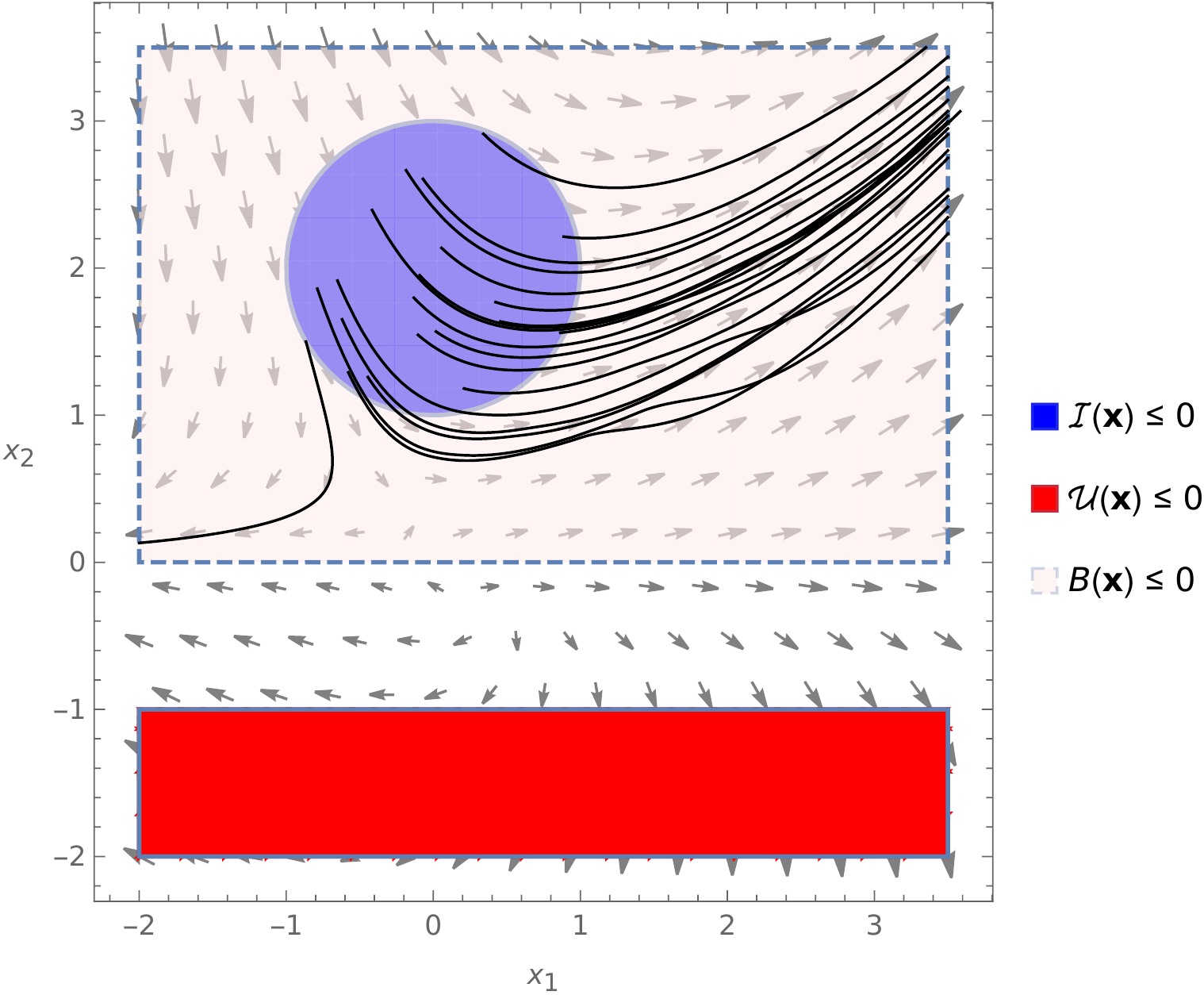}
	\end{center}
	\vspace*{-6mm}
	\caption{Phase portrait of the system in Example \ref{exmp:overview}. The arrows indicate the vector field and the solid curves are randomly sampled trajectories.}
	\vspace*{-5mm}
	\label{fig:overview}
\end{wrapfigure}

We show that the linearized program \eqref{eqn:bmiExpLinearized} 
is equivalent to an LMI optimization problem 
admitting polynomial-time algorithms, 
say the well-known \emph{interior-point methods} supported by most off-the-shelf SDP solvers.
Our iterative procedure starts with a strictly feasible initial solution $\zz^0$ 
to program \eqref{eqn:bmiExp} and terminates 
with $\lambda^2 \ge 0$ (subject to numerical round-off) and $a^2 = -0.00363421$, 
yielding the barrier certificate
{
	\setlength{\abovedisplayskip}{8pt}
	\setlength{\belowdisplayskip}{8pt}
\[
    B(\aaa^2, \xx) = -0.00363421 x_2 \leq 0. 
\]
}%
Fig.~\ref{fig:overview} depicts the system dynamics and the synthesized barrier certificate.

We remark that the aforementioned iterative procedure on solving a series of convex optimizations 
converges only to a local optimum of the BMI problem \eqref{eqn:bmiExp}. 
This means that, in some cases, it may miss the global optimum that induces a non-negative $\lambda^*$. 
We will present in Section~\ref{sec:bbframework} a solution to this problem 
by incorporating our iterative procedure into a branch-and-bound framework 
that searches for the global optimum in a divide-and-conquer fashion. 
%
%
%

\section{Mathematical Foundations}\label{sec:preliminaries}

\paragraph*{Notations.}
Let $\NN$, $\NN^+$, $\mathbb{R}$, $\RR^+$ and $\RR^+_0$ be respectively the set of natural, positive natural, real, positive real and non-negative real numbers. For a vector $\xx \in \RR^n$, $x_i$ refers to its $i$-th component and $\normsqrt{\xx}$ denotes the $\ell^2$-norm; for a matrix $A \in \mathbb{R}^{n \times m}$, $A(i, j)$ refers to its $(i, j)$-th element. Let $\mathbb{R}[\xx]$ be the polynomial ring in $\xx$ over the field $\mathbb{R}$. A polynomial $h \in \mathbb{R}[\xx]$ is \emph{sum-of-squares} (SOS) iff there exist polynomials $g_1, \ldots, g_k \in \mathbb{R}[\xx]$ such that $h = \sum_{i=1}^{k} g_i^2$. We denote by $\Sigma[\xx] \subset \mathbb{R}[\xx]$ the set of SOS polynomials over $\xx$. $\mathcal{S}^n$ denotes the space of $n \times n$ real, symmetric matrices. For $A \in \mathcal{S}^n$, $A \succeq 0$ means that $A$ is \emph{positive semidefinite} (psd, for short)\footnote{More generally, for $A, B \in \mathcal{S}^n$, $A \preceq B$ indicates that $B - A$ is positive semidefinite.}, i.e., $\forall \xx \in \mathbb{R}^n\colon \xx^\trans\! A \xx \ge 0$.
A matrix-valued function $\mathcal{B}\colon \mathbb{R}^n \to \mathcal{S}^m$ is \emph{psd-convex} on a convex set $\compactSet \subseteq \mathbb{R}^n$ if 
$
	\forall \xx_1, \xx_2 \in \compactSet \ldotp \forall \mu \in (0, 1)\colon
	\mathcal{B}(\mu \xx_1 + (1 - \mu) \xx_2) 
	\preceq \mu \mathcal{B}(\xx_1) + (1 - \mu) \mathcal{B}(\xx_2)
$.

\paragraph*{Differential Dynamical Systems.}
We consider a class of continuous dynamical systems modelled by ordinary differential equations of the autonomous type:
\begin{equation}
    \label{eqn:dynamics}
    \dot{\xx} = \ff(\xx)
\end{equation}%
where $\xx \in \RR^n$ is the \emph{state} vector, $\dot{\xx}$ denotes its temporal derivative ${\rm d}\xx/{\rm d}t$, with $t \in \RR^+_0$ modelling time, and $\ff\colon \RR^n \to \RR^n$ is a polynomial \emph{flow field} (or \emph{vector field}) that governs the evolution of the system. A polynomial vector field is local Lipschitz, and hence for some $T \in \RR^+ \cup \{\infty\}$, there exists a unique \emph{solution} (or \emph{trajectory}) $\sol_{\xx_0}\colon [0,T) \to \RR^n$ originating from any initial state $\xx_0 \in \RR^n$ such that (1) $\sol_{\xx_0}(0) = \xx_0$, and (2) $\forall \tau \in [0,T)\colon \frac{{\rm d}\sol_{\xx_0}}{{\rm d}t}\big\vert_{t=\tau} \!= \ff(\sol_{\xx_0}(\tau))$. We assume in the sequel that $T$ is the maximal instant up to which $\sol_{\xx_0}$ exists for all $\xx_0$.

\begin{remark}
Our techniques on synthesizing barrier certificates in this paper focus on differential dynamics of the form \eqref{eqn:dynamics}. However, we foresee no substantial difficulties in extending the results to multi-mode hybrid systems where extra constraints on the system evolution, e.g., guards, are present. 
\end{remark}

\paragraph*{Safety Verification Problem.}
Given a domain set $\domain \subseteq \RR^n$, an initial set $\init \subseteq \domain$ and an unsafe set $\unsafe \subseteq \domain$, the \emph{reachable set} of a dynamical system of the form \eqref{eqn:dynamics} at time instant $t \in [0,T)$ is defined as $\reach_{\init}(t) \define \{\sol_{\xx_0}(t) \mid \xx_0 \in \init\}$.
We denote by $\reach_{\init}$ the aggregated reachable set, i.e., the union of $\reach_{\init}(t)$ over $t \in [0,T)$\footnote{This subsumes the problem of unbounded-time safety verification where a unique solution exists over the infinite time horizon $[0,\infty)$.}. The system is said to be \emph{safe} iff $\reach_{\init} \cap \unsafe = \emptyset$, and \emph{unsafe} otherwise. For simplicity, we consider $\domain = \RR^n$ throughout this paper.

To avoid the explicit computation of the exact reachable set, which is usually intractable for nonlinear hybrid systems (cf., e.g., \cite{DBLP:journals/siglog/Fraenzle19}), barrier-certificate methods make use of a partial differential operator, termed the \emph{Lie derivative}, to capture the evolution of a barrier function along the vector field:
\begin{definition}[Lie derivative~\textnormal{\cite{kolar1993natural}}]
	Given a vector field $\ff\colon \RR^n \to \RR^n$ over $\xx$, the \emph{Lie derivative} of a polynomial function $B(\xx)$ along $\ff$, $\mathcal{L}_{\ff}^k B\colon \RR^n \to \RR$ of order $k \in \NN$, is
	\begin{equation*}
		\mathcal{L}_{\ff}^k B(\xx) \define
		\left\{
		\begin{array}{ll}
			B(\xx), \quad k=0,\\
			\left\langle\frac{\partial}{\partial \xx} \mathcal{L}_{\ff}^{k-1} B(\xx), \ff(\xx)\right\rangle, \quad k>0
		\end{array}
		\right.
	\end{equation*}%
	where $\langle\cdot,\cdot\rangle$ is the inner product of vectors, i.e., $\langle \uu, \vv\rangle \define \sum_{i=1}^{n}u_i v_i$ for $\uu, \vv \in \RR^n$.
\end{definition}
The Lie derivative $\mathcal{L}_{\ff}^k B(\xx)$ is essentially the $k$-th temporal derivative of the (barrier) function $B(\xx)$, and thus captures the change of $B(\xx)$ over time.

An \emph{inductive invariant} $\invt \subseteq \mathbb{R}^n$ of a dynamical system is a set of states such that all the trajectories starting from within $\invt$ remain in $\invt$:
\begin{definition}[Inductive invariant~\textnormal{\cite{PC08}}]\label{def:inv}
    Given a system \eqref{eqn:dynamics}, a set $\invt \subseteq \mathbb{R}^n$ is an \emph{inductive invariant} of system \eqref{eqn:dynamics} if and only if
    \begin{equation}
        \forall \xx_0 \in \invt \ldotp \forall t \in [0, T)\colon \sol_{\xx_0}(t) \in \invt.
    \end{equation}%
\end{definition}

In the sequel, we refer to inductive invariants simply as invariants. In~\cite{LZZ11}, a sufficient and necessary condition on being a polynomial invariant is proposed:
\begin{theorem}[Invariant condition~\textnormal{\cite{LZZ11}}]
    \label{thm:invariantCondition}
    Given a polynomial $B \in \mathbb{R}[\xx]$, its \emph{zero sub-level set} $\{ \xx \mid B(\xx) \leq 0 \}$ is an invariant 
    of system \eqref{eqn:dynamics} if and only if \footnote{In \eqref{eqn:invariantCondition}, $\bigwedge_{j = 0}^{i-1} \mathcal{L}_{\ff}^j B = 0$ is $\true$ for $i = 0$ by default. This applies in the sequel.}
    \begin{equation}
        \label{eqn:invariantCondition}
        B \leq 0 \implies 
        \bigvee\nolimits_{i = 0}^{\LieBound} \left(
        \left(\bigwedge\nolimits_{j = 0}^{i-1} \mathcal{L}_{\ff}^j B = 0\right) 
        \land \mathcal{L}_{\ff}^i B < 0\right)
        \lor \bigwedge\nolimits_{i = 0}^{\LieBound} \mathcal{L}_{\ff}^i B = 0
    \end{equation}%
    where $\LieBound \in \NN^+$ is a completeness threshold, i.e., a finite positive integer that bounds the order of Lie derivatives, 
    which can be computed using Gr\"{o}bner bases\footnote{$\LieBound$ is the minimal $i$ such that $\mathcal{L}_{\ff}^{i+1} B$ is in the polynomial ideal generated by $\mathcal{L}_{\ff}^0 B, \mathcal{L}_{\ff}^1 B, \ldots, \mathcal{L}_{\ff}^i B$. The ideal membership can be decided via Gr\"{o}bner basis.}.
\end{theorem}

In contrast, a \emph{barrier certificate} is a function whose zero sub-level set isolates an unsafe region $\unsafe$ from the reachable set $\reach_{\init}$ w.r.t.~some initial set $\init$:
\begin{definition}[Semantic barrier certificate~\textnormal{\cite{Platzer18FM}}]
\label{def:semanticBC}
Given a system \eqref{eqn:dynamics}, an initial set $\init$ and an unsafe set $\unsafe$, a \emph{barrier certificate} of \eqref{eqn:dynamics} is a differentiable function $B\colon \RR^n \to \RR$ satisfying
\begin{equation}
\label{eqn:semanticBc}
    \forall \xx_0 \in \init \ldotp \forall t \in [0,T)\colon B\left(\sol_{\xx_0}(t)\right) \leq 0 \quad \text{and} \quad 
    \forall \xx \in \unsafe\colon B(\xx) > 0.
\end{equation}%
\end{definition}
The existence of such a barrier certificate trivially implies safety of the system. Moreover, one may readily verify that if some set $\invt = \{ \xx \mid B(\xx) \leq 0 \}$ is an invariant and satisfies $(\init \subseteq \invt) \land (\invt \cap \unsafe = \emptyset)$, then $B(\xx)$ is a barrier certificate. 

As observed in \cite{Platzer18FM}, however, the semantic statement in Definition \ref{def:semanticBC} encodes merely the general \emph{principle of barrier certificates}~\cite{Gan17}, yet in itself is not that useful for safety verification because it explicitly involves the system solutions. Therefore, in order to enable efficient synthesis, the semantic condition on barrier certificates has been strengthened into a handful of different shapes (see, e.g., \cite{Prajna04, Kong13, yang2015exact, Gan17}, which all imply inductive invariance). It has been yet a long-standing challenge \emph{to find a barrier-certificate condition that is as weak as possible while admitting efficient synthesis algorithms}.


Our BMI encoding of the invariant barrier-certificate condition (cf.~Section~\ref{sec:formulation}) roots in Putinar's Positivstellensatz, which characterizes positivity of polynomials on a semi-algebraic set defined by a system of polynomial inequalities:

\begin{theorem}[Putinar's Positivstellensatz~\textnormal{\cite{lasserre2010moments}}]
	Let $\mathcal{K} = \{ \xx \mid \bigwedge_{i=1}^{m} g_i(\xx) \geq 0\}$ be a compact semi-algebraic set defined by $g_1, \ldots, g_m \in \RR[\xx]$.
        Assume the \emph{Archimedean condition} holds\footnote{This condition can be met by adding a (redundant) constraint $g_{m+1}(\xx) = L_0 - \norm{\xx} \leq 0$, provided that a bound $L_0 \in \RR^+$ is known such that $\forall \xx \in \mathcal{K}\colon L_0 - \norm{\xx} \geq 0$.}, i.e., there exists $L \in \RR^+$ such that $L - \norm{\xx} = \sigma_0(\xx) + \sum_{i=1}^{m} \sigma_i(\xx) g_i(\xx)$ for some $\sigma_0, \ldots, \sigma_m \in \Sigma[\xx]$.
	If $h \in \RR[\xx]$ is strictly positive on $\mathcal{K}$, then
	\begin{equation*}
		h(\xx) = \sigma_0(\xx) + \sum\nolimits_{i=1}^m \sigma_i(\xx) g_i(\xx)
	\end{equation*}%
	holds for some SOS polynomials $\sigma_0, \ldots, \sigma_m \in \Sigma[\xx]$.
\end{theorem}

We now recall a key technique used in our reduction to semidefinite optimizations. Given a symmetric matrix $X \in \mathcal{S}^n$ partitioned as
$X = 
\begin{pmatrix}
	A & C \\
	C^\trans & D
\end{pmatrix}$
with invertible $A$, the \emph{Schur complement} of $A$ in $X$ is defined as
$X / A \define D - C^\trans A^{-1} C$.
An important property of the Schur complement $X / A$ is that it characterizes the positive semidefiniteness of the block matrix $X$:
\begin{theorem}[Schur complement~\textnormal{\cite{boyd2004convex}}]
    \label{thm:schurComplement}
    If $A \succ 0$, then $X \succeq 0$ iff $X / A \succeq 0$. 
\end{theorem}%
We apply the Schur complement in Section~\ref{sec:algorithm} to transform nonlinear convex constraints into linear constraints.

\section{Invariant Barrier-Certificate Condition as BMIs}
\label{sec:formulation}
In this section, we present our \emph{invariant barrier-certificate condition} (see Definition~\ref{def:invBc}) based on the necessary and sufficient condition on being an inductive invariant (cf.~Theorem~\ref{thm:invariantCondition}), and show how to encode it as BMI constraints.

\subsection{Invariant Barrier-Certificate Condition}

\begin{definition}[Invariant barrier certificate]
    \label{def:invBc}
    Given a system \eqref{eqn:dynamics}, an initial set $\init$ and an unsafe set $\unsafe$, a polynomial function $B\colon \RR^n \to \RR$ is an \emph{invariant barrier certificate} of system \eqref{eqn:dynamics} if and only if
    \begin{enumerate}
        \item (initial): $\forall \xx \in \init\colon B(\xx) \leq 0$; 
        \item (consecution): $\forall \xx \in \RR^n\colon
                \bigwedge_{i = 1}^{\LieBound} \left(
                \left(\bigwedge_{j = 0}^{i-1} \mathcal{L}_{\ff}^j B(\xx) = 0\right)
                \implies \mathcal{L}_{\ff}^i B(\xx) \leq 0\right)$;
        \item (separation): $\forall \xx \in \unsafe\colon B(\xx) > 0$.
    \end{enumerate}
\end{definition}

Notice that the consecution constraint in Definition~\ref{def:invBc} involves Lie derivatives of orders up to $\LieBound \in \NN^+$, as is the case in Theorem~\ref{thm:invariantCondition}. Our invariant barrier-certificate condition hence generalizes existing conditions on barrier certificates, e.g., \cite{yang2015exact, zhang2018safety, CAV20BMI}, which consider Lie derivatives only up to the first order.

%
The consecution condition in Definition~\ref{def:invBc} is in fact equivalent to the invariant condition \eqref{eqn:invariantCondition} in Theorem~\ref{thm:invariantCondition} (cf.~Lemma~\ref{lem:eqConsecution} in Appendix~\ref{appendix_proofs}), thereby revealing the relation between an inductive invariant and an invariant barrier certificate:
\begin{theorem}[Inductive invariance]\label{thm:inductiveInvariance}
    Given a system \eqref{eqn:dynamics}, an initial set $\init$ and an unsafe set $\unsafe$. If $B(\xx)$ is an invariant barrier certificate, then $\invt = \{ \xx \mid B(\xx) \leq 0 \}$ is an invariant. Conversely, if $\invt = \{ \xx \mid B(\xx) \leq 0 \}$ is an invariant satisfying $\init \subseteq \invt$ and $\invt \cap \unsafe = \emptyset$, then $B(\xx)$ is an invariant barrier certificate.
\end{theorem}
%

It follows from Theorem~\ref{thm:inductiveInvariance} that our invariant barrier-certificate condition is the least conservative one on barrier certificates to attain inductive invariance.

\begin{remark}
We do not employ the invariant condition \eqref{eqn:invariantCondition} in Theorem~\refeq{thm:invariantCondition} as the constraint on the consecution of Lie derivatives. This is because our consecution condition in Definition~\refeq{def:invBc} is simpler, and in particular, amenable to more straightforward transformations to SOS constraints via Putinar's Positivstellensatz, as shown later in Subsection~\ref{subsec:BMIEncoding}.
\end{remark}

\begin{remark}
	For a fixed $0 < \mathfrak{N} < \LieBound$, the consecution condition in Definition~\ref{def:invBc} can be strengthened in the following way while preserving inductive invariance:
	\begin{multline*}
            \forall \xx \in \RR^n\colon
		\bigwedge\nolimits_{i = 1}^{\mathfrak{N}-1} \left(
		\left(\bigwedge\nolimits_{j = 0}^{i-1} \mathcal{L}_{\ff}^j B(\xx) = 0\right) 
		\implies \mathcal{L}_{\ff}^i B(\xx) \leq 0\right) \land \\
		\left(\left(\bigwedge\nolimits_{j = 0}^{\mathfrak{N}-1} \mathcal{L}_{\ff}^j B(\xx) = 0\right) 
		\implies \mathcal{L}_{\ff}^\mathfrak{N} B(\xx) < 0\right)
	\end{multline*}%
        where for the $\mathfrak{N}$-th Lie derivative, one needs $\mathcal{L}_{\ff}^\mathfrak{N} B(\xx) < 0$ (rather than $\mathcal{L}_{\ff}^\mathfrak{N} B(\xx) \leq 0$). In practice, using
        such a strengthened consecution condition ---with less sub-constraints to solve--- may yield more efficient synthesis.
\end{remark}
\subsection{Encoding as BMI Optimizations}
\label{subsec:BMIEncoding}

Next, we show how to encode synthesizing an invariant barrier certificate (cf.~Definition~\ref{def:invBc}) as an optimization problem subject to BMIs. To this end, we first recast the invariant barrier-certificate condition into a collection of SOS constraints\footnote{For simplicity, we assume that $\init$ and $\unsafe$ are both captured by a single polynomial. Our formulations, however, apply also to cases with basic semi-algebraic $\init$ or $\unsafe$.}. 
%

\begin{theorem}[Sufficient condition for invariant barrier certificate]
    \label{thm:invariantBcSosSufficient}
    Given a system \eqref{eqn:dynamics}, an initial set $\init = \{ \xx \mid \initBound(\xx) \leq 0 \}$ and an unsafe set $\unsafe = \{ \xx \mid \unsafeBound(\xx) \leq 0 \}$. A polynomial $B \in \mathbb{R}[\xx]$ is an invariant barrier certificate of \eqref{eqn:dynamics} if for some $\epsilon \in \RR^+$, there exist $v_{i, j} \in \mathbb{R}[\xx]$ and SOS polynomials $\sigma(\xx), \sigma'(\xx)$ s.t.
    \begin{enumerate}
        \item $-B(\xx) + \sigma(\xx) \initBound(\xx)$, 
        \item for all $1 \leq i \leq \LieBound$, $-\mathcal{L}_{\ff}^i B(\xx) + \sum\nolimits_{j = 0}^{i - 1} v_{i, j}(\xx) \mathcal{L}_{\ff}^j B(\xx)$, 
        \item $B(\xx)+ \sigma'(\xx) \unsafeBound(\xx) - \epsilon$
    \end{enumerate}
    are SOS polynomials.
\end{theorem}

By enforcing the Archimedean condition and applying Putinar's Positivstellensatz, we further derive a necessary condition of invariant barrier certificate:

\begin{theorem}[Necessary condition for invariant barrier certificate]
    \label{thm:invariantBcSosNecessary}
    \!Given a system \eqref{eqn:dynamics}, an initial set $\init = \{ \xx \mid \initBound(\xx) \leq 0 \}$ and an unsafe set $\unsafe = \{ \xx \mid \unsafeBound(\xx) \leq 0 \}$. If $B \in \mathbb{R}[\xx]$ is an invariant barrier certificate of \eqref{eqn:dynamics}, then for some $\epsilon \in \RR^+$, there exist $v_{i, j}  \in \mathbb{R}[\xx]$ and SOS polynomials $\sigma(\xx), \sigma'(\xx), \rho(\xx), \rho'(\xx), \rho_i''(\xx)$ s.t.~for any $L \in \RR^+$,
    \begin{enumerate}
        \item $-B(\xx) + \rho(\xx) (\normDisplay{\xx} - L) + \sigma(\xx) \initBound(\xx) + \epsilon$, 
        \item for all $1 \leq i \leq \LieBound$, $ -\mathcal{L}_{\ff}^i B(\xx) + \rho_i''(\xx) (\normDisplay{\xx} - L) + \sum\nolimits_{j = 0}^{i - 1} v_{i, j}(\xx) \mathcal{L}_{\ff}^j B(\xx) + \epsilon$, 
        \item $B(\xx) + \rho'(\xx) (\normDisplay{\xx} - L) + \sigma'(\xx) \unsafeBound(\xx)$
    \end{enumerate}
    are SOS polynomials. 
\end{theorem}

Notice that a polynomial $B(\xx)$ satisfying the sufficient condition in Theorem~\ref{thm:invariantBcSosSufficient} suffices as an invariant barrier certificate that witnesses safety of the system. In contrast, a polynomial $B(\xx)$ satisfying the necessary condition in Theorem~\ref{thm:invariantBcSosNecessary} may serve as a candidate invariant barrier certificate, and safety of the system can be concluded via a posterior check\footnote{Such a check inherits decidability of the first-order theory of real-closed fields~\cite{Tarski51}.}
of $B(\xx)$ per Definition~\ref{def:invBc}.
%

Next we show how to encode an SOS constraint of the shape ``$h(\xx) \in \Sigma[\xx]$'' in Theorems~\ref{thm:invariantBcSosSufficient} and~\ref{thm:invariantBcSosNecessary} as a BMI constraint. To this end, we first set a \emph{template polynomial}\footnote{A template polynomial $g(\aaa, \xx)$ is required to be linear in its parameters $\aaa$.} $B(\aaa, \xx)$ parameterized by unknown real coefficients $\aaa$ as the barrier certificate. We then proceed by setting templates for the remaining unknown polynomials (e.g., $v_{i, j}(\xx)$) and SOS polynomials (e.g., $\sigma(\xx)$ and $\rho(\xx)$) in $h(\xx)$, with all the parameters in these templates grouped into $\sss$. Observe that the parameterized SOS polynomial $h(\aaa,\sss,\xx)$ is of a bilinear form on the parameter spaces, i.e., $h(\aaa,\sss,\xx)$ is linear in $\aaa$ and $\sss$ separately. However, nonlinearity arises in the combined parameter space $(\aaa,\sss)$ due to the product couplings of $\aaa$ and $\sss$, i.e., $v_{i, j}(\sss_{i, j}, \xx) \mathcal{L}_{\ff}^j B(\aaa, \xx)$ in the consecution constraint.

Now the problem of synthesizing an invariant barrier certificate boils down to searching for an instantiation of the parameters $\aaa$ and $\sss$ such that the sufficient condition in Theorem~\ref{thm:invariantBcSosSufficient} holds (or alternatively, the necessary condition in Theorem~\ref{thm:invariantBcSosNecessary} holds and the posterior check passed). Such an instantiation of $\aaa$ (making $B(\aaa,\xx)$ an invariant barrier certificate) will be called \emph{valid} in the sequel.

Suppose that a parameterized SOS polynomial $h(\aaa,\sss,\xx)$ is of degree at most $2d$, with user-specified $d \in \NN$. Then $h(\aaa,\sss,\xx)$ can always be written in \emph{quadratic form} as $h(\aaa,\sss,\xx) = \mathbf{b}^\trans Q(\aaa,\sss) \mathbf{b}$, where $\mathbf{b} = (1, x_1, x_2, x_1 x_2, \ldots, x^d_n)$ is the \emph{basis vector} of size $p = \tbinom{n+d}{n}$ containing all monomials of degree up to $d$, and $Q(\aaa,\sss) \in \mathcal{S}^p$ is a parameterized real symmetric matrix known as the \emph{Gram matrix}~\cite{choi1995sums}\footnote{Extracting the Gram matrix amounts to solving a system of linear equations resulting from coefficient matching. The derived Gram matrix may contain extra unknowns if the system of linear equations admits multiple solutions, which nevertheless can be encoded in our subsequent workflow by enumerating the basis of its null space.}. An important fact states that $h(\aaa,\sss,\xx)$ is SOS if and only if $Q(\aaa,\sss) \succeq 0$.

Let $\mathcal{F}(\aaa, \sss) = - Q (\aaa,\sss)$. As per $h(\aaa,\sss,\xx)$, the matrix-valued function $\mathcal{F}(\aaa, \sss)$ is bilinear in $(\aaa,\sss)$. Observe that \emph{$h(\aaa,\sss,\xx)$ is SOS if and only if the BMI constraint $\mathcal{F}(\aaa, \sss) \preceq 0$ holds}. See Example~\ref{exmp:overview} for an illustration of this BMI encoding.

In general, $\mathcal{F}(\aaa, \sss)$ can be flattened in an expanded bilinear form as
\[
    \mathcal{F}(\aaa, \sss) = F + \sum\nolimits_{i=1}^m a_i H_i + \sum\nolimits_{j=1}^n s_j G_j + \sum\nolimits_{i=1}^m \sum\nolimits_{j=1}^n a_i s_j  F_{i, j}
\]%
where $m$ and $n$ are the size of $\aaa$ and $\sss$, respectively; $F, H_i, G_j, F_{i, j} \in \mathcal{S}^p$ are constant matrices.
Discharging the conditions of invariant barrier certificates hence amounts to solving the BMI feasibility problem of finding $\aaa$ and $\sss$ s.t.
\begin{equation}\label{eqn:bmiFeasiblity}
	\mathcal{F}_\iota(\aaa, \sss) \preceq 0, \quad \iota=1, 2, \ldots, l.
\end{equation}%
Here $\mathcal{F}(\aaa, \sss)$ is indexed by $\iota$ and $l$ is the number of SOS constraints involved.

To exploit well-developed techniques in optimization, the feasibility problem \eqref{eqn:bmiFeasiblity} is transformed to an optimization problem subject to BMI constraints:
\begin{maxi}
	{\lambda, \aaa, \sss}
	{\lambda}
	{\label{eqn:bmiBc}}
	{}
	\addConstraint{\mathcal{F}_\iota(\aaa, \sss) + \lambda I}{\preceq 0,\quad}{\iota=1, 2, \ldots, l.}
\end{maxi}%
A solution $(\lambda, \aaa, \sss)$ to~\eqref{eqn:bmiBc} is \emph{feasible} if it satisfies the BMIs in~\eqref{eqn:bmiBc}, and \emph{strictly feasible} if all the BMIs are satisfied with strict inequalities. We sometimes drop the $\lambda$ component in the solution when it is clear from the context. Notice that \emph{problem~\eqref{eqn:bmiFeasiblity} has a feasible solution if and only if the optimal value $\lambda^*$ in the BMI optimization problem~\eqref{eqn:bmiBc} is non-negative}.

To achieve (weak) completeness of our method in subsequent sections on solving the BMI optimization problem, we make the following assumption on the boundedness of the search space $(\aaa, \sss)$ of the optimization.

\begin{assumption}[Boundedness on the parameters]
    \label{ass:compactness}
    Every feasible solution $(\aaa, \sss)$ to the BMI problem~\eqref{eqn:bmiBc} is in a compact set with non-empty interior, i.e.,
    \begin{equation*}
        (\aaa, \sss) \in \compactSet_{\aaa} \times \compactSet_{\sss} = \left\{(\aaa,\sss) \mathrel{\big|} \normDisplay{\aaa} \leq L_{\aaa}, \normDisplay{\sss} \leq L_{\sss}\right\}
    \end{equation*} 
    for some known bounds $L_{\aaa}, L_{\sss} \in \RR^+$. 
\end{assumption}

\begin{remark}
	The boundedness on $\aaa$ in Assumption~\ref{ass:compactness} makes sense in practice since we usually prefer barrier certificates with bounded coefficients. Moreover, when the bilinear functions $\mathcal{F}_\iota(\aaa, \sss)$ in~\eqref{eqn:bmiBc} are affine in $\aaa$ and $\sss$, i.e., with a zero constant matrix $F$, the parameters $\aaa$ and $\sss$ can be scaled independently by any positive factor. Therefore in this case, w.l.o.g, one may simply set $L_{\aaa} = L_{\sss} = 1$.
\end{remark}

\section{Solving BMI Optimizations via DCP}
\label{sec:algorithm}

The BMI optimization problem~\eqref{eqn:bmiBc}, derived from the synthesis problem, is known to be NP-hard and contains non-convex constraints~\cite{toker1995np}, and hence is not amenable to efficient (polynomial-time) algorithms committed to solving convex optimizations. In this section, we present an algorithm for solving general BMI optimizations via difference-of-convex programming~\cite{tao1986algorithms,le2018dc}, which solves a series of convex sub-problems that approaches a local optimum of~\eqref{eqn:bmiBc}.

For brevity, we consider optimization problems with a single BMI constraint\footnote{Multiple BMI constraints can be joined as a single BMI in a block-diagonal fashion.}:
\begin{maxi}
	{\zz=(\xx, \yy)}
	{g(\zz)}
	{\label{eqn:bmip}}
	{}
	\addConstraint{}{\mathcal{B}(\xx, \yy) \define F + \sum_{i=1}^m x_i H_i + \sum_{j=1}^n y_j G_j + \sum_{i=1}^m \sum_{j=1}^n x_i y_j  F_{i, j}}{\preceq 0}
\end{maxi}%
where the objective function $g\colon \RR^{m+n} \to \RR$ is linear in $\zz = (\xx$, $\yy)$; $F, H_i, G_j, F_{i, j} \in \mathcal{S}^p$ are constant symmetric matrices.

\subsection{Difference-of-Convex Decomposition}
The key challenge in solving the BMI problem \eqref{eqn:bmip} is its non-convexity, that is, the matrix-valued function $\mathcal{B}(\xx, \yy)$ is, in general, not psd-convex.

There have been attempts, most pertinently in~\cite{dinh2011combining}, to decompose a bilinear \mbox{function} as a difference between two psd-convex functions, known as the \emph{difference-of-convex} (DC) \emph{decomposition}, such that the optimization in its decomposed form enjoys well-established techniques in difference-of-convex programming~\cite{tao1986algorithms,le2018dc}. The DC decomposition in~\cite{dinh2011combining}, however, is confined to BMIs of a specific structure, namely, $X^\trans Y + Y^\trans X \preceq 0$, where $X$ and $Y$ are matrix variables containing variables $x_i$ and $y_j$, respectively. The more general bilinear function $\mathcal{B}(\xx, \yy)$ in \eqref{eqn:bmip} does unfortunately not admit straightforward forms of decomposition such as those in~\cite[Lemma~3.1]{dinh2011combining}.

In what follows, we present a difference-of-convex decomposition of the matrix-valued function $\mathcal{B}(\xx, \yy)$, inspired by~\cite{wang2016feasibility}, using eigendecomposition.

First, observe that the function $\mathcal{B}(\xx, \yy)$ can be written as 
\begin{equation}
    \label{eqn:bmiKronecker}
    \mathcal{B}(\xx, \yy) = 
    \begin{pmatrix}
        \xx \otimes I \\
        \yy \otimes I
    \end{pmatrix}^\trans
    \begin{pmatrix}
        0 & \Gamma \\
        \Gamma^\trans & 0
    \end{pmatrix}
    \begin{pmatrix}
        \xx \otimes I \\
        \yy \otimes I
    \end{pmatrix} 
    \mathbin{+} 
    \begin{pmatrix}
        \Omega_1 & \Omega_2 
    \end{pmatrix} 
    \begin{pmatrix}
        \xx \otimes I \\
        \yy \otimes I
    \end{pmatrix}
    + F
\end{equation}
where $\otimes$ denotes the Kronecker product: for two matrices $A \in \mathbb{R}^{a \times b}$ and $B \in \mathbb{R}^{c \times d}$, $A \otimes B \define [A(1,1) B, \ldots, A(1,b) B; \text{\rotatebox[origin=c]{-15}{$\cdots$}}; A(a,1) B, \ldots, A(a,b) B]  \in \mathbb{R}^{a c \times b d}$, $0$ represents the zero matrices with compatible dimensions, and
\begin{equation*}
    \label{eqn:defGammaOmega}
    \Gamma = \frac{1}{2} 
    \begin{pmatrix}
        F_{1, 1} & \dots & F_{1, n} \\
        \vdots & \ddots & \vdots \\
        F_{m, 1} & \dots & F_{m, n}
    \end{pmatrix},\quad
    \Omega_1 = 
    \begin{pmatrix}
        H_1 & \dots & H_m
    \end{pmatrix},\quad
    \Omega_2 = 
    \begin{pmatrix}
        G_1 & \dots & G_n
    \end{pmatrix}.
\end{equation*}%

The form of \eqref{eqn:bmiKronecker} implies that 
$\mathcal{B}(\xx, \yy)$ is psd-convex if the matrix
$
    M = \begin{pmatrix}
            0 & \Gamma \\
            \Gamma^\trans & 0
        \end{pmatrix}
$
is positive semidefinite. Unfortunately, as \cite[Theorem~1]{wang2016feasibility} points out, 
for a non-trivial bilinear function $\mathcal{B}(\xx, \yy)$, 
$M$ may not be positive semidefinite.

Nevertheless, the matrix $M$ can always be decomposed as $M = M_1 - M_2$ with $M_1, M_2 \succeq 0$, i.e., a difference between two psd-matrices. 
One way to do so is to use the \emph{eigendecomposition} of the (real symmetric\footnote{$M$ thus only has real eigenvalues.}) matrix $M \in \mathcal{S}^{(m+n)p}$. That is,
$
    M = V^\trans D V
$,
where the orthogonal matrix $V$ contains the eigenvectors of $M$; $D$ is a diagonal matrix whose diagonal elements are the eigenvalues of $M$.

Let $D^+$ be the matrix obtained by setting all negative elements of $D$ to zero and $D^- = D^+ - D$. We have 
\begin{equation*}
    M = \underbrace{V^\trans D^+ V}_{M_1} - \underbrace{V^\trans D^- V}_{M_2}. 
\end{equation*}%
It follows that $M_1, M_2 \succeq 0$ and therefore we find a DC decomposition of $\mathcal{B}(\xx, \yy)$:

\begin{theorem}[Difference-of-convex decomposition]
    \label{thm:psdConvexity}
    The following form
    \begin{equation}
    	\label{eqn:dc}
    	\mathcal{B}(\xx, \yy) = 
    	\mathcal{B}^+(\xx, \yy) - \mathcal{B}^-(\xx, \yy)
    \end{equation}%
    where
    \vspace*{-.2cm}
    \begin{equation*}
    	\begin{split}
    		\mathcal{B}^+(\xx, \yy) &=     
    		\begin{pmatrix}
    			\xx \otimes I \\
    			\yy \otimes I
    		\end{pmatrix}^\trans
    		M_1
    		\begin{pmatrix}
    			\xx \otimes I \\
    			\yy \otimes I
    		\end{pmatrix}
    		\mathbin{+} 
    		\begin{pmatrix}
    			\Omega_1 & \Omega_2 
    		\end{pmatrix} 
    		\begin{pmatrix}
    			\xx \otimes I \\
    			\yy \otimes I
    		\end{pmatrix}
    		+ F \\
    		\mathcal{B}^-(\xx, \yy) &= 
    		\begin{pmatrix}
    			\xx \otimes I \\
    			\yy \otimes I
    		\end{pmatrix}^\trans
    		M_2
    		\begin{pmatrix}
    			\xx \otimes I \\
    			\yy \otimes I
    		\end{pmatrix}
    	\end{split}
    \end{equation*}%
	is a difference-of-convex decomposition of $\mathcal{B}(\xx, \yy)$. Namely, the matrix-valued functions $\mathcal{B}^+(\xx, \yy)$ and $\mathcal{B}^-(\xx, \yy)$ are psd-convex on $\RR^{m+n}$.
\end{theorem}

\begin{remark}
	In practice, the aforementioned matrices $M$, $M_1$ and $M_2$ induced by eigendecomposition are often highly sparse. One can hence exploit the sparsity to improve the algorithmic performance of the DCP-based synthesis approach.
\end{remark}


\subsection{Reduction to LMIs}
On top of the DC decomposition (cf.~Theorem~\ref{thm:psdConvexity}), we can now apply a standard iterative procedure in difference-of-convex programming~\cite{dinh2011combining} to solve the BMIs.

The core idea of the procedure is to iteratively solve a series of convex sub-problems. More specifically, given a feasible solution $\zz^k = (\xx^k, \yy^k)$ to the BMI optimization problem~\eqref{eqn:bmip}, the ``concave part'' $-\mathcal{B}^-(\xx, \yy)$ in \eqref{eqn:dc} is linearized around $\zz^k$, thereby yielding a series of convex programs ($k = 0, 1, \ldots$):

\begin{maxi}
	{\zz=(\xx, \yy)}
	{g(\zz) + \frac{1}{2} \delta \normDisplay{\zz - \zz^k}}
	{\label{eqn:bmipLinearized}}
	{}
	\addConstraint{}{\mathcal{B}^+(\zz) - \mathcal{B}^-\left(\zz^k\right) - \mathcal{DB}^-\left(\zz^k\right)\left(\zz -\zz^k\right)}{\preceq 0}
\end{maxi}%
where $\mathcal{DB}^-(\zz)\colon \RR^{m+n} \to \mathcal{S}^p$ is the derivative of the matrix-valued function $\mathcal{B}^-$ at $\zz$, i.e., a linear mapping from a vector $\uu \in \RR^{m+n}$ to a matrix in $\mathcal{S}^p$:
\begin{equation*}
	\mathcal{DB}^-(\zz) (\uu) \define \sum\nolimits_{i = 1}^{n+m} u_i \frac{\partial \mathcal{B}^-}{\partial z_i}(\zz).
\end{equation*}%
An extra regularization term $\frac{1}{2} \delta \norm{\zz - \zz^k}$ with $\delta < 0$ is added in~\eqref{eqn:bmipLinearized} to enforce that $g(\zz)$ strictly increases after each iteration until it stabilizes, which can be encoded as a second-order cone constraint and embedded in SDP solving.

Note that the linearized problem~\eqref{eqn:bmipLinearized} is convex and therefore can be solved efficiently\footnote{The global optimum of~\eqref{eqn:bmipLinearized} is attainable under standard assumptions, e.g., Slater's condition and the second-order sufficient KKT conditions~\cite{boyd2004convex}.} via methods including, among others, augmented Lagrangian methods~\cite{li2018qsdpnal} and gradient descent methods~\cite{boyd2004convex}. Furthermore, the Schur complement in Theorem~\ref{thm:schurComplement} implies that~\eqref{eqn:bmipLinearized} can be reformulated as an LMI problem:

\begin{theorem}
    \label{thm:bmiToLmi}
    The quadratic matrix inequality (QMI) constraint
    \begin{equation*}
       \mathcal{B}^+(\zz) - \mathcal{B}^-\left(\zz^k\right) - \mathcal{DB}^-\left(\zz^k\right)\left(\zz -\zz^k\right) \preceq 0
    \end{equation*} 
    in \eqref{eqn:bmipLinearized} is equivalent to the LMI constraint\footnote{This transforms a QMI with matrices in $\mathcal{S}^{p}$ to an LMI with matrices in $\mathcal{S}^{(m+n+1)p}$.}
    \begin{equation*}
    	\begingroup 
    	\setlength\arraycolsep{3pt}
        \begin{pmatrix}
            -I \quad & N (\zz \otimes I) \\
            (\zz \otimes I)^\trans N^\trans \quad  & - \mathcal{B}^-\left(\zz^k\right) - \mathcal{DB}^-\left(\zz^k\right)\left(\zz -\zz^k\right) + \Omega (\zz \otimes I) + F
        \end{pmatrix}
    	\endgroup
        \preceq 0 
    \end{equation*}
    where $N$ is the square root matrix of $M_1$, i.e., $M_1$ = $N^\trans N$, and $\Omega = \begin{pmatrix} \Omega_1 & \Omega_2 \end{pmatrix}$.  
  \end{theorem}

Theorem~\ref{thm:bmiToLmi} entails that the series of linearized convex sub-problems of the form~\eqref{eqn:bmipLinearized} can be solved alternatively by most off-the-shelf SDP solvers designated for discharging LMIs via polynomial-time algorithms, say the interior-point methods.
Furthermore, by taking the optimum of the $k$-th sub-problem to be the next linearization point $\zz^{k+1}$, we obtain an iterative procedure for solving general BMIs, as depicted in Algorithm~\ref{alg:localBMI}.

\begin{algorithm}[t]
	\caption{\toolname{BMI-DC}: Solving BMIs based on DC decomposition}
	\label{alg:localBMI}
	\SetKwInput{Input}{input}\SetKwInOut{Output}{output}\SetNoFillComment
	\Input{A BMI optimization problem~\eqref{eqn:bmip} with a strictly feasible initial solution $\zz^0$.}
	\Output{A sequence of feasible solutions $S = \left\{\zz^0, \ldots, \zz^k\right\}$ to the BMI optimization.}
    $k \gets 0$;~
    $S \gets \left\{ \zz^0 \right\}$\;
    $M \gets \text{reformulation of~\eqref{eqn:bmip} as~\eqref{eqn:bmiKronecker}}$\;
    $(M_1, M_2) \gets \text{DC decomposition of } M$ as in~\eqref{eqn:dc}\;
    \Repeat{$\normDisplaysqrt{\zz^k - \zz^{k-1}} < \varepsilon~\text{\upshape for a given tolerance } \varepsilon \in \RR^+_0$}{
    	 Construct the convex sub-problem~\eqref{eqn:bmipLinearized} out of $(M_1, M_2)$ linearized around $\zz^k$\;
         $\zz^{k+1} \gets \text{optimum of the program~\eqref{eqn:bmipLinearized}}$\;
         $S \gets S \cup \left\{ \zz^{k+1} \right\}$\tcp*{$S\;\mathtt{keeps\;track\;of\;visited\;points}$}
         $k \gets k+1$\;
    }
	\Return $S$\;
\end{algorithm}

Algorithm~\ref{alg:localBMI} falls into the DCP framework~\cite{dinh2011combining} and thus enjoys useful properties, e.g., soundness, termination and convergence as follows.

\begin{theorem}[Soundness]
    \label{thm:soundnessLocal}
    Every solution $\zz^i  = (\xx^i, \yy^i) \in S$ with $i = 0, \ldots, k$ returned by Algorithm~\ref{alg:localBMI} is a feasible solution to the original BMI problem~\eqref{eqn:bmip}.
\end{theorem}

The result below states termination and convergence of Algorithm~\refeq{alg:localBMI}  in terms of \emph{KKT points} of~\eqref{eqn:bmip}, i.e., solutions fulfilling the 
KKT conditions~\cite{boyd2004convex} of~\eqref{eqn:bmip}\footnote{Addressing the KKT conditions in detail falls outside the scope of this paper.}.
%

\begin{theorem}[Termination and convergence]
	\label{thm:convergenceLocal}
	If~\eqref{eqn:bmip} has finitely many KKT points, then (1) for $\varepsilon \in \RR^+$, Algorithm~\refeq{alg:localBMI} terminates; (2) for $\varepsilon = 0$, Algorithm~\refeq{alg:localBMI} visits an infinite sequence of solutions converging to a KKT point.
\end{theorem}


We remark that, under some sufficient KKT conditions and regularity conditions~\cite{boyd2004convex}, a KKT point suffices as a local optimum. In this case, the infinite sequence $\{\zz^i\}_{i \in \NN}$ of points visited by Algorithm~\refeq{alg:localBMI} (for $\varepsilon = 0$) converges to a  local optimum of~\eqref{eqn:bmip}.

\subsection{Finding the Initial Solution}\label{subsec:initialSolution}

The iterative procedure in Algorithm~\refeq{alg:localBMI} starts with a fed-by-oracle strictly feasible initial solution $\zz^0$ to the BMI problem~\eqref{eqn:bmip}. Finding such an initial solution, however, is non-trivial in general due to the non-convexity of~\eqref{eqn:bmip}. We argue though, that a strictly feasible initial solution can be obtained for the BMI problem of the form~\eqref{eqn:bmiBc} induced by the barrier-certificate synthesis problem.

Recall that in the BMI problem~\eqref{eqn:bmiBc}, bilinearity arises from the multiplication of 
$B(\aaa, \xx)$ with some unknown multiplier polynomials parameterized by $\sss$. One way to reduce the BMI constraints to LMIs is to fix every multiplier polynomial to be a non-negative constant, thereby yielding a linear program:

\begin{maxi}
	{\lambda, \aaa}
	{\lambda}
	{\label{eqn:lmiExponetialBc}}
	{}
	\addConstraint{\mathcal{F}_\iota(\aaa, \sss)\big\vert_{\sss = \left(c_\iota, 0, \ldots, 0\right)} + \lambda I}{\preceq 0,\quad}{\iota=1, 2, \ldots, l}
\end{maxi}%
where $\sss$ in $\mathcal{F}_\iota(\aaa, \sss)$ is substituted by $(c_\iota, 0, \ldots, 0)$ with $c_\iota \in \RR^+_0$, which encodes a non-negative constant multiplier polynomial. Observe that no $\sss$-variable is involved in~\eqref{eqn:lmiExponetialBc} and the constraints therein are linear in $\aaa$.

Apparently, a strictly feasible solution $(\lambda, \aaa)$ to~\eqref{eqn:lmiExponetialBc} induces a strictly feasible solution $(\lambda, \aaa, (c_\iota, 0, \ldots, 0))$ to~\eqref{eqn:bmiBc} as well. Moreover, we have
\begin{lemma}\label{lem:LMIsolution}
    The LMI program~\eqref{eqn:lmiExponetialBc} always has a strictly feasible solution.
\end{lemma}

As a consequence, a strictly feasible solution to the BMI problem~\eqref{eqn:bmiBc} can be obtained by solving the LMI problem~\eqref{eqn:lmiExponetialBc}. In fact, when considering Lie derivatives only up to the first order, solving (the feasibility counterpart of)~\eqref{eqn:lmiExponetialBc} is exactly the procedure to synthesize either an \emph{exponential barrier certificate}~\cite{Kong13} (with $c_\iota \in \RR^+$) or a \emph{convex barrier certificate}~\cite{Prajna04} (with $c_\iota = 0$). Algorithm~\refeq{alg:localBMI} therefore subsumes existing synthesis techniques in the sense that any valid barrier certificate synthesized by methods in~\cite{Kong13,Prajna04} can also be discovered by Algorithm~\refeq{alg:localBMI}. Moreover, an alternative way to reduce the BMI constraints to LMIs is to fix the multipliers to be some given non-trivial (SOS) polynomials~\cite{zeng2016darboux}.

\begin{remark}
	Different choices of the multiplier constants $c_\iota$ in~\eqref{eqn:lmiExponetialBc} may lead to different initial solutions fed to Algorithm~\refeq{alg:localBMI}, thereby considerably different number of iterations until termination. In practice, techniques like randomization are worth exploring when choosing these multiplier constants.
\end{remark}


\section{Incorporating in a Branch-and-Bound Framework}
\label{sec:bbframework}

The aforementioned iterative procedure on solving a series of convex optimizations converges only to a local optimum of the BMI problem~\eqref{eqn:bmiBc} (or more generally,~\eqref{eqn:bmip}). This means that, in some cases, it may miss the global optimum that induces a non-negative $\lambda^*$. We present in this section a solution to this problem by incorporating the iterative procedure into a branch-and-bound framework that searches for the global optimum in a divide-and-conquer fashion, as is a common technique in non-convex optimizations.
%

The basic idea is as follows. We first try to solve the BMI problem~\eqref{eqn:bmiBc}
by Algorithm~\ref{alg:localBMI} over the compact parameter space $(\compactSet_{\aaa}, \compactSet_{\sss})$. If a valid solution, (i.e, a solution that contains a valid parameter $\bar{\aaa} \in \compactSet_{\aaa}$ such that $B(\bar{\aaa}, \xx)$ is an invariant barrier certificate) is found, then the corresponding barrier certificate can be obtained. Otherwise, we keep bisecting $\compactSet_{\aaa}$ and apply Algorithm~\ref{alg:localBMI} over each bisection\footnote{The validity of $\bar{\aaa} \in \compactSet_{\aaa}$ does not depend on $\sss$, thus we do not partition $\compactSet_{\sss}$.}. The procedure, as depicted in Algorithm~\ref{alg:bbAlgorithm} in a recursive manner, terminates when a valid parameter is found or the partition is fine enough.

\begin{algorithm}[t]
\caption{\toolname{Branch-and-Bound}: Searching for a valid parameter $\bar{\aaa}$}
\label{alg:bbAlgorithm}
\SetKwInput{Input}{input}\SetKwInOut{Output}{output}\SetNoFillComment
\Input{A BMI optimization problem of the form~\eqref{eqn:bmiBc} with $\compactSet_{\aaa} = \{\aaa \mid \norm{\aaa} \leq L_{\aaa}\}$.}
\Output{A valid parameter $\bar{\aaa}$, or otherwise $\bot$ indicating a failure.}
    \lIf(\tcp*[f]{$\mathtt{abort\;on\;fine\mbox{-}enough\;partitions\;(}\eta \in \RR^+\mathtt{)}$}){$L_{\aaa} < \eta$}{\Return $\bot$}
    \tcc{$\mathtt{sample\mbox{-}and\mbox{-}check\;is\;not\;necessary\;if\;Theorem\;\ref{thm:invariantBcSosNecessary}\;is\;used}$}
    $\bar{\aaa} \gets \text{a randomly-sampled point in } \compactSet_{\aaa}$\;\label{lin:startSample}
    \lIf(\tcp*[f]{$\mathtt{check\;validity\;(inductive\;invariance)}$}){$\bar{\aaa}$ \text{\upshape is valid}}{\Return $\bar{\aaa}$\label{lin:endSample}}
    \If(\tcp*[f]{$S_{\mathit{glb}}\;\mathtt{contains\;a\;global\;set\;of\;visited\;points}$}){$\textit{proj}_{\aaa}(S_{\textit{glb}}) \cap \compactSet_{\aaa} = \emptyset$\label{lin:checkS}}
    {
        $S \gets \text{apply } \toolname{BMI-DC}$ in Algorithm~\ref{alg:localBMI} \text{to}~\eqref{eqn:bmiBc} with initial solution in $(\compactSet_{\aaa}, \compactSet_{\sss})$\;
        $S_{\textit{glb}} \gets S_{\textit{glb}} \cup S$\;
        \tcc{$\mathtt{checking\;validity\;is\;not\;necessary\;if\;Theorem\;\ref{thm:invariantBcSosSufficient}\;is\;used}$}\label{lin:comment}
        \lIf{\text{\upshape a valid parameter} $\bar{\aaa} \in \textit{proj}_{\aaa}(S)$ \text{\upshape is found}}{\Return $\bar{\aaa}$}\label{lin:checkForNec}
    }
    $(\compactSet_{\aaa}^1, \compactSet_{\aaa}^2) \gets \textit{bisect}(\compactSet_{\aaa})$\tcp*{$\mathtt{partition\;the\;parameter\;space}$}\label{lin:startPartition}
    $\bar{\aaa} \gets \toolname{Branch-and-Bound}(\compactSet_{\aaa}^1)$\;
    \lIf{$\bar{\aaa} \neq \bot$}{\Return $\bar{\aaa}$}
    \lElse{\Return $\toolname{Branch-and-Bound}(\compactSet_{\aaa}^2)$}\label{lin:endPartition}
\end{algorithm}

Algorithm~\ref{alg:bbAlgorithm} takes as input a BMI problem of the form~\eqref{eqn:bmiBc} that encodes either the sufficient condition in Theorem~\ref{thm:invariantBcSosSufficient} or the necessary condition in Theorem~\ref{thm:invariantBcSosNecessary} for invariant barrier certificates. In the former case, a sample-and-check process (Line~\ref{lin:startSample}--\ref{lin:endSample}) is necessary to attain (weak) completeness (see Theorem~\ref{thm:bbCompleteness}). The conditional statement in Line~\ref{lin:checkS} rules out parameter (sub-)spaces that have already been explored, which is the case when the projection of some visited point in $S_{\textit{glb}}$ (a global set that keeps track of visited points by Algorithm~\ref{alg:localBMI}, initialized as $\emptyset$) onto $\aaa$ is in the current parameter space.

The following theorem claims a weak completeness result: our method guarantees to find a barrier certificate when there exists an inductive invariant (in the form of a given template) that suffices to certify safety of the system.
%

\begin{theorem}[Weak Completeness]
    \label{thm:bbCompleteness}
    Algorithm~\refeq{alg:bbAlgorithm} returns a valid parameter $\bar{\aaa} \in \compactSet_{\aaa}$, if (1) the partition granularity is fine enough (i.e., small enough $\eta \in \RR^+$), (2) the degrees of multiplier polynomials and SOS polynomials used to form~\eqref{eqn:bmiBc} are large enough, and (3) there exists, for the given template $B(\aaa,\xx)$, a strictly valid parameter $\hat{\aaa} \in \compactSet_{\aaa}$ (i.e., any parameter in some neighborhood of $\hat{\aaa}$ is valid).
\end{theorem}
%

\begin{remark}
	The bisection operation in Algorithm~\ref{alg:bbAlgorithm} induces ---in the worst case--- an exponential blow-up in the number of branches. In practice, one can prune branches inducing only negative objective values, via, e.g., convex relaxation~\cite{kheirandishfard2018convex}. 
\end{remark}

\section{Experimental Results}\label{sec:experiments}

We have carried out a prototypical implementation\footnote{Available at~\faGithub~\url{https://github.com/Chenms404/BMI-DC}.} of our synthesis techniques in Wolfram \textsc{Mathematica}, which was selected due to its built-in primitives for SDP, polynomial algebra and matrix operations.  Given a safety verification problem as input, our implementation works toward discovering an invariant barrier certificate (in the form of a given template) that witnesses unbounded-time safety of the system. A collection of benchmark examples (detailed in Appendix~\ref{appendix_examples}) has been evaluated on a 2.10GHz Intel Xeon processor with 376GB RAM running 64-bit CentOS Linux 7.

\begin{table}[h]
	\caption{Empirical results on benchmark examples (time in seconds)} 
	\label{tab:results}
	\begin{center}
		\adjustbox{width=1\textwidth}{
			\begin{tabular}{lcccc>{\centering}m{1mm}rcrc>{\centering}m{1mm}rc>{\centering}m{1mm}rc}
				\toprule
				\multirow{2}{*}{Example name} & &
				\multirow{2}{*}{$n_{\mathsf{sys}}$} &
				\multirow{2}{*}{$d_{\mathsf{flow}}$} &
				\multirow{2}{*}{$d_{\mathsf{BC}}$} & &
				\multicolumn{4}{c}{\toolname{BMI-DC}}& &
				\multicolumn{2}{c}{\toolname{PENLAB}}&  & 
				\multicolumn{2}{c}{\toolname{SOSTOOLS}} \\
				\cmidrule(lr){7-10} \cmidrule(lr){12-13} \cmidrule(lr){15-16}
				& & & & & & \#iter. & & time & verified & & time~ & verified & & time & verified \\
				\midrule	
				\expname{overview}~\cite{djaballah2017construction}        &   &  2   &  2        & 1   & &   2 &  & \textbf{0.03}  & \cmark & & 0.31  &  \cmark& & 0.07     &      \cmark      \\ 	
				\expname{contrived}          &  &2   &  1        & 2   &  &  0& & \textbf{0.01}  &  \cmark & & 0.48     &     \cmark    &  &   0.75     &    \cmark      \\ 
				\expname{lie-der}~\cite{LZZ11}        &   &  2   &  2        & 1  & &     0 & & \textbf{0.01}   &    \cmark   &    & 0.22        &    \cmark   &  &  0.04    &        \cmark        \\ 
				\expname{lorenz}~\cite{djaballah2017construction}      &     &  3   &  2        & 2  &  &    8 & & \textbf{2.37}  &     \cmark   & & 75.11 &   \xmark   &  &   1.47     &     \xmark          \\ 
				\expname{lti-stable}~\cite{DBLP:conf/cav/GaoKDRSAK19}    &       &  2   &  1        & 2  &   &   0 & & \textbf{0.01}      &   \cmark &   & 0.23               &   \cmark  &   &    0.14    &      \cmark         \\ 
				\expname{lotka-volterra}~\cite{goubault2014finding}       &    &  3   &  2        & 1  &  &    3 & & \textbf{0.07}      &   \cmark  & & 0.36   &    \cmark  &  &    0.21    &       \cmark        \\ 
				
				\expname{clock}~\cite{RatschanS05}      &     &  2   &  3        & 1 &   &    0 & & \textbf{0.01}       &   \cmark   &   & 0.88  &   \xmark   & &    0.18    &        \xmark        \\ 
				\expname{lyapunov}~\cite{ratschan2010providing}              &   &  3   &  3        & 2 &  &     4 & & 1.25       &    \cmark   &   & 56.98  &   \xmark  &  &   \textbf{0.35}     &        \cmark        \\ 
				\expname{arch1}~\cite{sogokon2016non}       &          &  2   &  5        & 2   & &    0 & & \textbf{0.01}      &    \cmark   &   & 33.76  &   \xmark  &   &   0.31     &      \cmark         \\ 
				\expname{arch2}~\cite{sogokon2016non}       &          &  2   &  2        & 2   & &   5 & & \textbf{0.37}       &     \cmark   &  & 0.38  &   \xmark  &   &     0.17   &       \xmark        \\ 
				\expname{arch3}~\cite{sogokon2016non}        &         &  2   &  3        & 2   &  &   1 & & \textbf{0.07}   &    \cmark    &      & 0.54  &  \cmark &  &     0.18   &        \cmark          \\ 
				\expname{arch4}~\cite{sogokon2016non}        &         &  2   &  2        & 1    &  &  2 & & 0.09       &    \cmark   &  & 0.49 &   \xmark  &  &    \textbf{0.06}    &        \cmark        \\ 
				
				\expname{barr-cert1}~\cite{Prajna04}     &      &  2   &  3        & 2 &   &   12  & & \textbf{0.85}   &     \cmark  &  & 2.53    &   \xmark  &  &    0.09    &        \xmark        \\ 
				\expname{barr-cert2}~\cite{djaballah2017construction}     &       &  2   &  2        & 2 &  &     6 & & \textbf{1.57}    &    \cmark  &   & 1.16       &    \xmark &   &    0.15    &      \cmark         \\ 
				\expname{barr-cert3}~\cite{zhang2018safety}        &         &  2   &  2        & 1 &  &     0 & & \textbf{0.01}       &   \cmark    &  & 0.20 &   \cmark &  &    0.11    &         \xmark        \\ 
				\expname{barr-cert4}~\cite{zhang2018safety}        &         &  2   &  3        & 2  & &     13 & & \textbf{0.96}     &      \cmark  &    & 0.89  &  \xmark &  &   0.23     &        \xmark          \\ 	
				
				\expname{fitzhugh-nagumo}~\cite{DBLP:conf/cdc/SassiGS14}        &        &  2  &  3        & 2 &   &    2 & & \textbf{0.16}      &      \cmark  &    &1.24 &  \cmark   &  &    0.25    &     \xmark           \\ 	
				\expname{stabilization}~\cite{DBLP:conf/hybrid/SassiS15}        &        &  3  &  2        & 2 &   &    9 & & 2.88     &      \cmark    &  & 55.22 &   \cmark   & &    \textbf{0.11}    &      \cmark          \\ 
				\expname{lie-high-order}       &   &  2   &  1        & 2  & &     32 & & \textbf{4.12}   &    \cmark   &    &1.56        &   \xmark   &  & 0.25    &       \xmark        \\ 
				\expname{raychaudhuri}~\cite{ferragut2015seeking}        &        &  4  &  2        & 2 &   &    34 & & \textbf{9.51}      &      \cmark &     & 33.64 &   \xmark  &  &   0.14     &      \xmark          \\ 
				
				\expname{focus}~\cite{ratschan2006constraints}         &       &  2  &  1        & 4  & &    100 & & 54.89      &      \xmark    &  & 0.95 &   \xmark  &  &     0.48   &        \xmark        \\ 
				
				\expname{sys-bio1}~\cite{klipp2008systems}         &        &  7   &  2        & 2 &   &    2& &  73.22    &     \namark   &     & 101.95 &  \namark  &  &   1.35     &      \namark           \\ 
				\expname{sys-bio2}~\cite{klipp2008systems}       &         &  9   &  2        & 1  & &     1&  & 1.03       &    \namark  &  & 15.54 &    \namark  &  &      0.16  &       \namark        \\ 
				\expname{quadcopter}~\cite{DBLP:conf/cav/GaoKDRSAK19}       &         &  12  &  1        & 1  &  &    0 & & 0.03      &      \namark   &   & 65.42 &   \namark  &  &   0.36     &     \namark           \\ 
				
				\bottomrule
			\end{tabular}
		}
	\end{center}
	\vspace*{-2mm}
	\scriptsize{$n_{\mathsf{sys}}$: system dimension;  $d_{\mathsf{flow}}$: maximal flow-field degree; $d_{\mathsf{BC}}$: degree of the template barrier certificate.\\
		\#iter.: number of iterations. 0 means that the initial solution (cf.~Subsection~\ref{subsec:initialSolution}) is valid.\\
		verified: the synthesized barrier certificate is valid (\cmark), invalid (\xmark) or inconclusive (\namark, beyond the capability of quantifier elimination in \textsc{Mathematica} and nonlinear reasoning in \textsc{Z3}).\\
		time: CPU-time, excluding that for casting the BMIs/LMIs. Boldface marks the winner among \cmark's.
	}
\end{table}

Table~\ref{tab:results} reports the empirical results. \toolname{BMI-DC} concerns our locally-convergent Algorithm~\ref{alg:localBMI} for solving BMIs (encoding the sufficient condition in Theorem~\ref{thm:invariantBcSosSufficient}) based on DC decomposition. We compare our approach with \toolname{PENLAB}~\cite{fiala2013penlab} ---an off-the-shelf solver in \textsc{Matlab} for directly discharging the same BMI problems (with no guarantee on convergence)--- and \toolname{SOSTOOLS}~\cite{sostools2013} ---for solving LMIs derived from Prajna and Jadbabaie's original barrier-certificate condition~\cite{Prajna04}. The comparison is performed under the same problem configurations\footnote{For \toolname{PENLAB} and \toolname{SOSTOOLS}, we use their optimized, built-in criteria for termination and methods for finding the initial solutions.}. Due to numerical errors caused by floating-point computations and the fact that reaching the local/global optimum does not necessarily yield a valid barrier certificate, we additionally perform a posterior check, via both the quantifier-elimination procedure in \textsc{Mathematica} and the SMT solver \textsc{Z3}~\cite{z3}, of the synthesized candidate barrier certificate per Definition~\ref{def:invBc}.

Table~\ref{tab:results} shows that \toolname{BMI-DC} suffices to synthesize valid barrier certificates in most of the examples within a reasonable number of iterations (i.e., the number of convex sub-problems solved by SDP). This however does not cover all the cases: for the \expname{focus} example, the solution is close enough to a local optimum (after 100 iterations) but yields still an invalid barrier certificate. This problem can be solved (if there exists an invariant barrier certificate as specified) by enforcing the branch-and-bound framework as presented in Section~\ref{sec:bbframework}. The phase portraits of a selected set of examples and the synthesized invariant barrier certificates are depicted in Fig.~\ref{fig:visualization} (see more in Appendix~\ref{appendix_examples}).

\begin{figure}[t]
	\centering
	\resizebox{\textwidth}{!}{
		\begin{tabular}{ccc}
			\subfloat[\expname{lti-stable}]{~~~~\includegraphics[scale=0.28]{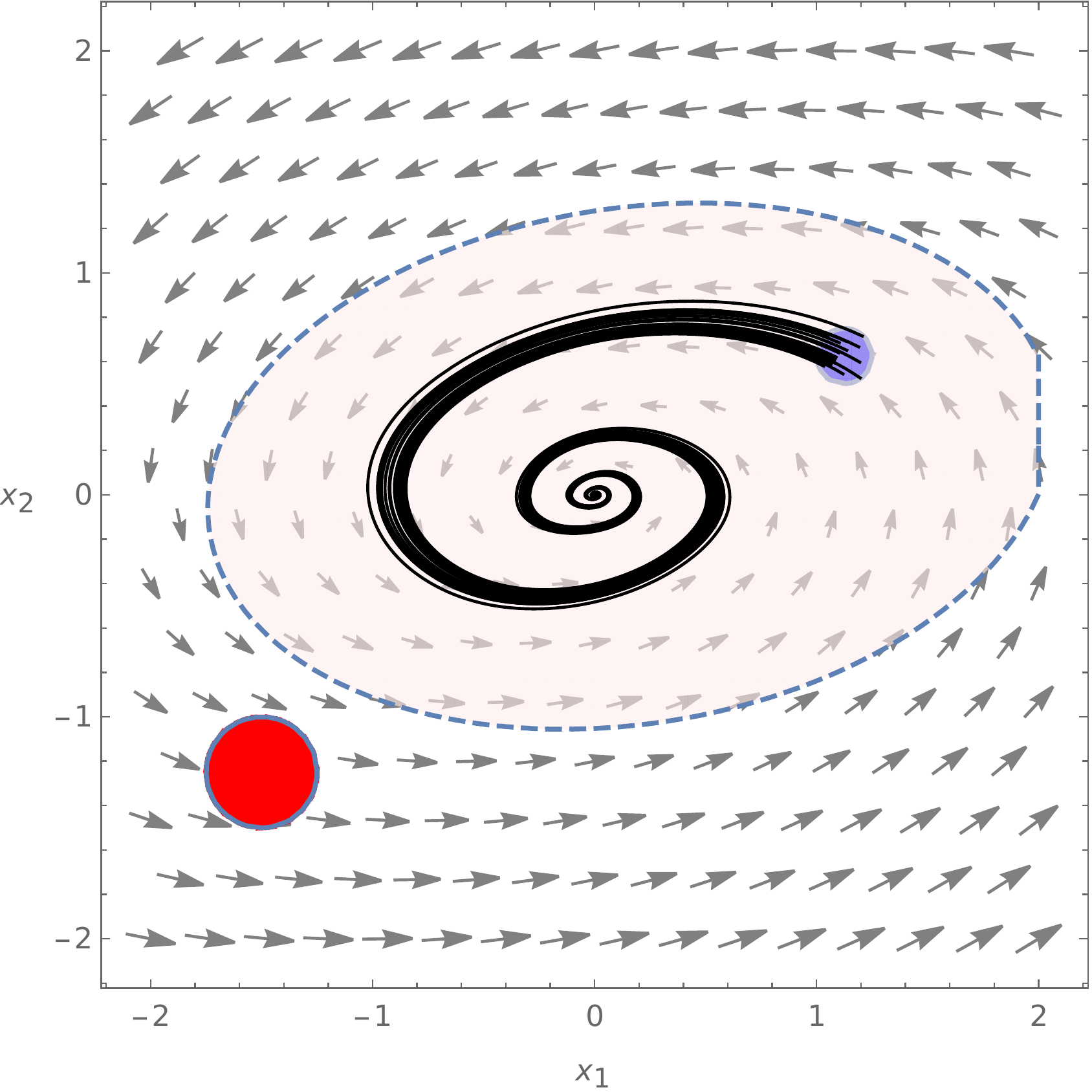}~~~~\label{fig:lti-stable}}&
			\hspace*{.1cm}
			\subfloat[\expname{lyapunov}]{~~~~\includegraphics[scale=0.28]{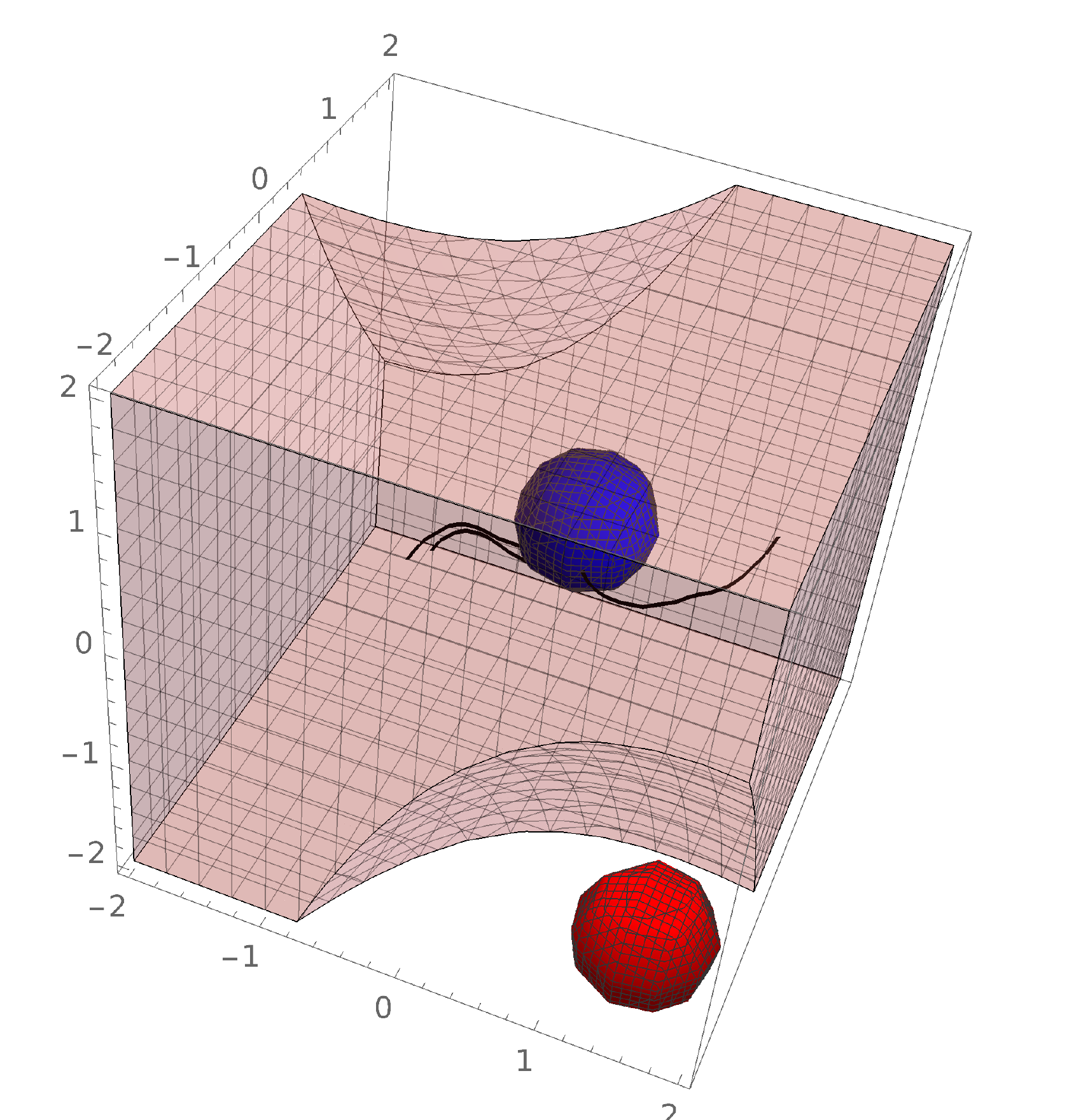}~~~~\label{fig:C2}}
			\hspace*{-.1cm}& 
			\subfloat[\expname{barr-cert2}]{~~~~\includegraphics[scale=0.382]{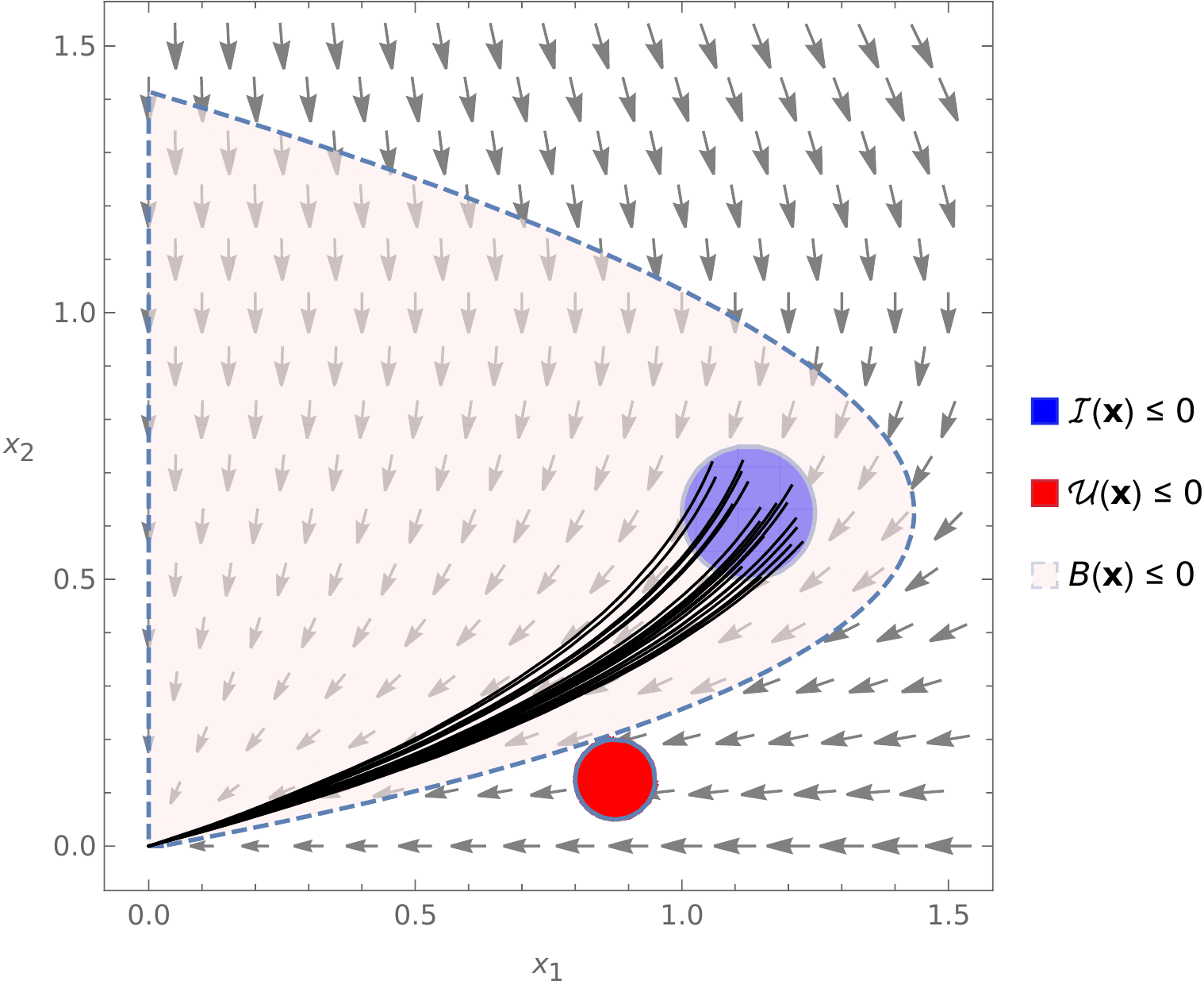}~~~~\label{fig:exmp3}}
		\end{tabular}
	}
	\caption{Phase portraits of a selected set of examples with the synthesized invariant barrier certificates. The arrows indicate the vector field (hidden in 3D-graphics for a clear presentation) and the solid curves are randomly sampled trajectories.}\label{fig:visualization}
\end{figure}

The comparison in Table~\ref{tab:results} suggests that (1) Our invariant barrier-certificate condition recognizes more barrier certificates than the original (more conservative) condition as implemented in \toolname{SOSTOOLS}. In particular, the \expname{lie-high-order} example does admit an inductive invariant in the form of the given template, but none of the existing barrier-certificate conditions~\cite{yang2015exact, zhang2018safety, CAV20BMI} ---concerning Lie derivatives only up to the first order--- recognizes it, since we have $\mathcal{L}_{\ff}^1 B (\xx) = 0$ for some $\xx$ on the boundary of $B$ and hence it requires to exploit the second-order Lie derivative $\mathcal{L}_{\ff}^2 B$; (2) Our DCP-based synthesis algorithm finds more barrier certificates in less time than directly solving the BMI problems via non-convex optimization techniques as implemented in \toolname{PENLAB}.

We remark that symbolic methods based on, e.g., quantifier elimination~\cite{LZZ11}, can hardly deal with any of the examples listed in Table~\ref{tab:results} due to the prohibitively high computation complexity. Moreover, it would be desirable to pursue a comparison with the augmented Lagrangian method for solving BMIs as proposed in~\cite{CAV20BMI}, which unfortunately is not yet possible due to the unavailability of the implementation thereof. We will discuss crucial differences to~\cite{CAV20BMI} in Section~\ref{sec:related}.
\section{Related Work}\label{sec:related}

As surveyed in~\cite{DBLP:journals/siglog/Fraenzle19}, the research community has, over the past three decades, extensively addressed the automatic verification of safety-critical hybrid systems. The almost universal undecidability of the unbounded-time reachability problem~\cite{ACH95}, however, confines the sound key-press routines to either semi-decision procedures or approximation schemes, most of which address bounded-time verification by, e.g., computing the finite-time image of a set of initial states.

Invariant generation\cite{Kapur17}, amongst others, is a well-established approximation scheme that provides a reliable witness for safety (or equivalently, unreachability) of dynamical systems over the infinite time horizon. Invariants can be constructed in various forms, e.g., 
barrier certificates~\cite{Prajna04,Platzer18FM} and differential invariants~\cite{PC08,LZZ11}. With a priori specified templates, the invariant synthesis problem can be reduced to numerical optimizations or constraint solving, as in, e.g.,~\cite{tiwari2003approximate,SSM04a,gulwani2008constraint,kapinski2014simulation}. 

Most pertinently, Prajna and Jadbabaie proposed in their seminal work~\cite{Prajna04} a concept coined \emph{barrier certificate} to encode invariants. To enable efficient synthesis via semidefinite programming, the barrier-certificate condition in~\cite{Prajna04} strengthens the general condition encoding inductive invariance. Since then, significant efforts have been investigated in developing more relaxed (i.e., weaker) forms of barrier-certificate condition that still admit efficient synthesis, thereby leading to, e.g., exponential-type barrier certificates~\cite{Kong13}, Darboux-type barrier certificates~\cite{zeng2016darboux}, general barrier certificates~\cite{Gan17} and vector barrier certificates~\cite{Platzer18FM}. To attain efficient synthesis, these barrier-certificate conditions share a common property on convexity.
That is, if for some $\aaa_1, \aaa_2 \in \mathbb{R}^m$, $B(\aaa_1, \xx)$ and $B(\aaa_2, \xx)$ both satisfy the barrier-certificate condition, then for any $0 < \mu < 1$, $B(\mu \aaa_1 + (1 - \mu) \aaa_2, \xx)$ must also satisfy the barrier-certificate condition.

However, neither the semantic barrier-certificate condition~\eqref{eqn:semanticBc} encoding the general principle of barrier certificates~\cite{Platzer18FM,Gan17} nor the inductive invariant condition~\eqref{eqn:invariantCondition} is convex. This means, when resorting to convex barrier-certificate conditions, one may miss some potential barrier certificates that suffice as inductive invariants witnessing safety. Therefore, non-convex conditions were suggested~\cite{yang2015exact}, for which the synthesis problem can be reduced to BMI problems solvable via customized schemes, e.g., the augmented Lagrangian method~\cite{CAV20BMI} and the alternating minimization algorithm~\cite{zhang2018safety}. Our synthesis techniques also exploit a BMI reduction, with three crucial differences: (1) our invariant barrier-certificate condition is equivalent to the inductive invariant condition in the sense of Theorem~\ref{thm:inductiveInvariance}, and thus is less conservative than all the aforementioned conditions which consider Lie derivatives only up to the first order; (2) our DCP-based techniques for solving BMIs naturally inherit appealing results on convergence and (weak) completeness, which are not (and can hardly be) provided by the approaches in~\cite{yang2015exact,CAV20BMI,zhang2018safety}; (3) our DCP-based iterative procedure visits only feasible solutions to the original BMI problem, and hence whenever a solution that induces a non-negative objective value is found, we can safely terminate the algorithm and claim a feasible solution to the original BMI problem, which may yield a valid barrier certificate. This is not the case for the approaches in~\cite{yang2015exact,CAV20BMI,zhang2018safety}.

Beyond barrier certificates, Wang and Rajamani~\cite{wang2016feasibility} investigated the feasibility problem of general BMI problems with an application to multi-objective nonlinear observer design. The technique of eigendecomposition was also used therein to conduct the DC decomposition. The decomposed concave part, however, is simply ignored and no iterative procedure that exhibits convergence to a local optimum can be provided.

The idea of augmenting a locally-convergent algorithm with a branch-and-bound framework to find the global optimum has been exploited in the realm of optimization~\cite{goh1995global} and control~\cite{tuan2000new}. In contrast, our method is designed for the specific problem of barrier-certificate synthesis, and hence our branch-and-bound algorithm concerns only the parameter space of $\aaa$, i.e., coefficients of the template barrier certificate.

Finally, we refer interested readers to other approaches to solving BMI problems,
e.g., rank minimization~\cite{ibaraki2001rank,orsi2006newton,recht2010guaranteed}, sequential SDP~\cite{correa2004global,eggers2012improving}, as well as methods committed to general non-convex optimizations, e.g., interior point trust-region~\cite{dennis1998trust,leibfritz2002interior,chiu2016method}, successive linearization~\cite{kanzow2005successive} and primal-dual interior point~\cite{yamashita2012local}.

\section{Conclusion}\label{sec:conclusion}
Barrier certificates are powerful tools to prove time-unbounded safety of hybrid systems. 
We have presented a new condition on barrier certificates ---the invariant barrier-certificate condition. This condition is by far the least conservative one on barrier certificates, and can be shown as the weakest possible one to attain inductive invariance. We showed that our invariant barrier-certificate condition can be reformulated as an optimization problem subject to bilinear matrix inequalities, which can be solved by our locally-convergent algorithm based on difference-of-convex programming. By incorporating this algorithm into a branch-and-bound framework, we obtained a weak completeness result. 
Experiments on benchmark examples suggested that our invariant barrier-certificate condition recognizes more barrier certificates than existing conditions, and that our DCP-based algorithm is more efficient than directly solving the BMIs via off-the-shelf solvers. 

We stress that our techniques for solving BMIs are of a general nature rather than being confined to barrier-certificate synthesis. Interesting future directions include to extend our method to other synthesis problems, e.g., discovering invariants and/or termination proofs of deterministic/probabilistic programs. 


\paragraph*{\bf Acknowledgements.}
The authors would like to thank Hengjun Zhao for the fruitful discussion on differential dynamics requiring high-order Lie derivatives. 

%
%
\bibliographystyle{splncs04}
\bibliography{ref.bib}

\newpage 

\setcounter{section}{0}

\begin{subappendices}
	\renewcommand{\thesection}{\Alph{section}}

\section{Proofs of Lemmas and Theorems}\label{appendix_proofs}



\begin{lemma}[Equivalence of Lie consecution]\label{lem:eqConsecution}
	The consecution condition in Definition~\refeq{def:invBc} holds 
        if and only if the invariant condition \eqref{eqn:invariantCondition} 
        in Theorem~\refeq{thm:invariantCondition} holds.
\end{lemma}

\begin{proof}
	We prove both the ``if'' and the ``only if'' part by contradiction.
	
	For the ``if'' part, suppose that the invariant condition \eqref{eqn:invariantCondition} holds but the consecution condition is invalid. The latter implies that for some $\xx_0 \in \RR^n$ and $1 \leq i_0 \leq \LieBound$,
	\begin{equation}
		\label{eqn:invBcCase}
		\left(\bigwedge\nolimits_{j = 0}^{i_0 - 1} \mathcal{L}_{\ff}^j (\xx_0) = 0\right) 
		\land \mathcal{L}_{\ff}^{i_0} B(\xx_0) > 0.
	\end{equation}%
	Note that \eqref{eqn:invBcCase} implies $B(\xx_0) = 0$. From \eqref{eqn:invariantCondition}, it follows that either
	\begin{equation}
		\label{eqn:invariantConditionCase1}
		\bigwedge\nolimits_{i = 0}^{\LieBound} \mathcal{L}_{\ff}^i B(\xx_0) = 0
	\end{equation}%
	holds, or there exists $0 \leq i_1 \leq \LieBound$ such that
	\begin{equation}
		\label{eqn:invariantConditionCase2}
		\left(\bigwedge\nolimits_{j = 0}^{i_1-1} \mathcal{L}_{\ff}^j B(\xx_0) = 0\right) 
		\land \mathcal{L}_{\ff}^{i_1} B(\xx_0) < 0
	\end{equation}%
	holds. However, 
	\begin{itemize}
		\item \eqref{eqn:invariantConditionCase1} cannot hold as $\mathcal{L}_{\ff}^{i_0}B(\xx_0) = 0$ in \eqref{eqn:invariantConditionCase1} but $\mathcal{L}_{\ff}^{i_0}B(\xx_0) > 0$ in \eqref{eqn:invBcCase};
		\item for $i_1 \leq i_0$, \eqref{eqn:invariantConditionCase2} cannot hold as $\mathcal{L}_{\ff}^{i_1} B(\xx_0) < 0$ in \eqref{eqn:invariantConditionCase2} but $\mathcal{L}_{\ff}^{i_1} B(\xx_0) \geq 0$ in \eqref{eqn:invBcCase}; 
		\item for $i_1 > i_0$, \eqref{eqn:invariantConditionCase2} cannot hold as $\mathcal{L}_{\ff}^{i_0} B(\xx_0) = 0$ in \eqref{eqn:invariantConditionCase2} but $\mathcal{L}_{\ff}^{i_0} B(\xx_0) > 0$ in \eqref{eqn:invBcCase}. 
	\end{itemize}
	%
	
	For the ``only if'' direction, suppose that the consecution condition in Definition~\ref{def:invBc} holds but the invariant condition \eqref{eqn:invariantCondition} is invalid. The latter implies that there exists $\xx_0$ such that $B(\xx_0) \leq 0$ and
	\begin{equation}
		\label{eqn:invariantConditionNegCase}
		\neg\left(\left(\bigwedge\nolimits_{j = 0}^{i-1} \mathcal{L}_{\ff}^j B (\xx_0) = 0\right) 
		\land \mathcal{L}_{\ff}^i B (\xx_0) < 0\right)
	\end{equation}%
	holds for any $0 \leq i \leq \LieBound$.
	
	For $i = 0$, \eqref{eqn:invariantConditionNegCase} yields that $B(\xx_0) \geq 0$. Together with the premise $B(\xx_0) \leq 0$, we have $B(\xx_0) = \mathcal{L}_{\ff}^{0} B(\xx_0) = 0$. 
	Now, by taking the case $i = 1$ in the consecution condition, we deduce $\mathcal{L}_{\ff}^{1} B(\xx_0) \leq 0$. Meanwhile, for $i = 1$, \eqref{eqn:invariantConditionNegCase} yields $\mathcal{L}_{\ff}^{1} B(\xx_0) \geq 0$. It thus follows that $\mathcal{L}_{\ff}^{1} B(\xx_0) = 0$. 
	Analogously, by taking $i = 2, \ldots, \LieBound$, we conclude $\mathcal{L}_{\ff}^{i} B(\xx_0) = 0$ for all $0 \leq i \leq \LieBound$. This is exactly encoded in \eqref{eqn:invariantCondition} (the rightmost conjunctive clause) and hence contradicts the assumption that \eqref{eqn:invariantCondition} is invalid. Therefore, the consecution condition implies \eqref{eqn:invariantCondition}.
	\qed
\end{proof}

\begin{proof}[of Theorem~\ref{thm:inductiveInvariance}]
	The claim is an immediate consequence of Lemma~\ref{lem:eqConsecution}.
   \qed
\end{proof}

\begin{proof}[of Theorem~\ref{thm:invariantBcSosSufficient}]
	It can be shown that the $k$-th condition in Theorem~\ref{thm:invariantBcSosSufficient} implies the $k$-th condition in Definition~\ref{def:invBc}, for $k =1, 2, 3$. For instance, the first condition in Theorem~\ref{thm:invariantBcSosSufficient} requires that $-B(\xx) + \sigma(\xx) \initBound(\xx)$ is an SOS polynomial (and thus non-negative), we therefore have $B(\xx) \leq \sigma(\xx) \initBound(\xx)$. It follows that $\forall \xx \in \{ \xx \mid \initBound(\xx) \leq 0 \}\colon B(\xx) \leq 0$. A similar argument applies to the other two conditions.
	\qed
\end{proof}

\begin{proof}[of Theorem~\ref{thm:invariantBcSosNecessary}]
	The invariant barrier-certificate condition in Definition~\ref{def:invBc} characterizes positivity of polynomials over certain sets. By adding a ``ball'' constraint $\norm{\xx} - L \leq 0$ to those sets (thus achieving the Archimedean condition), we can apply Putinar's Positivstellensatz to rewrite those polynomials into SOS forms. 
	
	For instance, the initial condition in Definition~\ref{def:invBc} implies that $-B(\xx) + \epsilon$ is strictly positive on $\mathcal{K} = \{ \xx \mid -\initBound(\xx) \geq 0 \land - (\norm{\xx} - L) \geq 0 \}$. Putinar's Positivstellensatz can then be applied to show that $- B(\xx) + \epsilon = \sigma_0(\xx) -  \rho(\xx) (\norm{\xx} - L) - \sigma(\xx) \initBound(\xx)$ holds for some SOS polynomials $\sigma_0(\xx)$, $\rho(\xx)$ and $\sigma(\xx)$. The first condition in Theorem~\ref{thm:invariantBcSosNecessary} then follows immediately.
	A similar argument applies to the other two conditions.
	\qed
\end{proof}

\begin{proof}[of Theorem~\ref{thm:psdConvexity}]
	We first show the psd-convexity of $\mathcal{B}^+(\xx, \yy)$. Let $\zz = (\xx, \yy) \in \RR^{m+n}$.
	According to~\cite[Proposition 1]{DBLP:journals/mp/Shapiro97}, $\mathcal{B}^+(\zz) = \mathcal{B}^+(\xx, \yy)$ is psd-convex if (and only if) for any $\vv \in \RR^{p}$, the function
	$
	\phi_{\vv}(\zz) = \vv^\trans \mathcal{B}^+(\zz) \vv
	$
	is convex. Note that
	\begin{equation*}
		\begin{split}
			\phi_{\vv}(\zz) 
			&= \vv^\trans 
			\begin{pmatrix}
				\zz \otimes I
			\end{pmatrix}^\trans
			M_1
			\begin{pmatrix}
				\zz \otimes I
			\end{pmatrix}
			\vv
			+
			\vv^\trans
			\begin{pmatrix}
				\Omega_1 & \Omega_2
			\end{pmatrix}
			\begin{pmatrix}
				\zz \otimes I
			\end{pmatrix}
			\vv
			+
			\vv^\trans F \vv  \\
			&= (\zz \otimes \vv)^\trans
			M_1
			(\zz \otimes \vv)
			+
			\vv^\trans
			\begin{pmatrix}
				\Omega_1 & \Omega_2
			\end{pmatrix}
			(\zz \otimes \vv)
			+
			\vv^\trans F \vv.
		\end{split}
	\end{equation*}%
	Then, for any $\mu_1 \in (0, 1)$ and $\mu_2 = 1 - \mu_1$, we have, for any $\zz_1, \zz_2 \in \RR^{m+n}$,
	\begin{align*}
		&\phi_{\vv}(\mu_1 \zz_1 + \mu_2 \zz_2) - 
		(\mu_1 \phi_{\vv}(\zz_1) + \mu_2 \phi_{\vv}(\zz_2)) \\
		=\ &
		(\mu_1 (\zz_1 \otimes \vv) + \mu_2 (\zz_2 \otimes \vv))^\trans M_1 
		(\mu_1 (\zz_1 \otimes \vv) + \mu_2 (\zz_2 \otimes \vv)) \\
		&- 
		\mu_1 (\zz_1 \otimes \vv)^\trans M_1 (\zz_1 \otimes \vv)
		- \mu_2 (\zz_2 \otimes \vv)^\trans M_1 (\zz_2 \otimes \vv) \\
		=\ &
		\mu_1 \mu_2 (\zz_2 \otimes \vv)^\trans M_1 (\zz_1 \otimes \vv) + 
		\mu_1 \mu_2 (\zz_1 \otimes \vv)^\trans M_1 (\zz_2 \otimes \vv) \\
		&- 
		\mu_1 \mu_2 (\zz_1 \otimes \vv)^\trans M_1 (\zz_1 \otimes \vv) - 
		\mu_1 \mu_2 (\zz_1 \otimes \vv)^\trans M_1 (\zz_1 \otimes \vv) \\
		=\ &
		- \mu_1 \mu_2 ((\zz_1 - \zz_2) \otimes \vv)^\trans M_1 
		((\zz_1 - \zz_2) \otimes \vv) \\
                \leq\ & 0 \tag{positive semidefiniteness of $M_1$} 
	\end{align*}%
	which means that $\phi_{\vv}(\zz)$ is convex. Thus, $\mathcal{B}^+(\xx, \yy)$ is psd-convex.
	
	The psd-convexity of $\mathcal{B}^-(\xx, \yy)$ can be shown in an analogous way. 
	\qed
\end{proof}

\begin{proof}[of Theorem~\ref{thm:bmiToLmi}]
	Note that the square root matrix $N$ of $M_1$ exists since $M_1 \succeq 0$\footnote{As we have $M_1 = V^\trans D^+ V$ (with only non-negative eigenvalues in $D^+$) from the eigendecomposition of $M$, the matrix $N$ can be computed as $N = V^\trans (D^+)^{1/2} V$, where $(D^+)^{1/2}$ is the diagonal matrix whose diagonal elements are square roots of those in $D^+$.}. 
	The claim then follows immediately by applying the Schur complement in Theorem~\ref{thm:schurComplement}. 
	\qed
\end{proof}

\begin{proof}[of Theorem~\ref{thm:soundnessLocal}]
	We prove by induction on $i$.
        The base case holds as $\zz^0$ is assumed to be a feasible solution to~\eqref{eqn:bmip}.
        For the induction step, we show that $\zz^{i+1}$ is a feasible solution 
        to~\eqref{eqn:bmip} if $\zz^i$ is a feasible solution to~\eqref{eqn:bmip}. 
        Since $\zz^{i+1}$ is a feasible solution to~\eqref{eqn:bmipLinearized} linearized at $\zz^i$, 
        it suffices to show that the feasible set of~\eqref{eqn:bmipLinearized} 
        is a subset of the feasible set of~\eqref{eqn:bmip}. 
	
	Theorem~\ref{thm:psdConvexity} shows that $\mathcal{B}^-(\zz)$ is psd-convex, then by~\cite[Lemma~2.2~(b)]{dinh2011combining}, we have 
	\begin{equation}\label{eq:soundnessProof1}
		\mathcal{B}^-(\zz) - \mathcal{B}^-\left(\zz^i\right) \succeq \mathcal{DB}^-\left(\zz^i\right)\left(\zz - \zz^i\right).
	\end{equation}%
	In the meantime, $\zz^i$ is a feasible solution to~\eqref{eqn:bmipLinearized} and thus fulfils
	\begin{equation}\label{eq:soundnessProof2}
		\mathcal{B}^+(\zz) - \mathcal{B}^-\left(\zz^i\right) - \mathcal{DB}^-\left(\zz^i\right)\left(\zz - \zz^i\right) \preceq 0.
	\end{equation}%

	Combining~\eqref{eq:soundnessProof1} and~\eqref{eq:soundnessProof2}, we have
	$
	\mathcal{B}(\xx, \yy) = 
	\mathcal{B}^+(\zz) - \mathcal{B}^-(\zz) \preceq 0
	$
	which is exactly the BMI constraint of~\eqref{eqn:bmip}. This completes the proof.
	\qed
\end{proof}

\begin{proof}[of Theorem~\ref{thm:convergenceLocal}]
	Let $\bar{S} = \{\zz^i\}_{i \in \NN}$ be the infinite sequence of visited points for $\varepsilon = 0$.
	
	We first show that (2) implies (1).
	Assume that (2) holds, i.e., $\bar{S}$ converges (to a KKT point of~\eqref{eqn:bmip}), then by Cauchy's criterion for convergence, we have $\forall \varepsilon \in \RR^+ \ldotp \exists k \in \NN^+\colon \normsqrt{\zz^k - \zz^{k-1}} < \varepsilon$ (with $\zz^k, \zz^{k-1} \in \bar{S}$). Algorithm~\refeq{alg:localBMI} thus terminates.
	
	It then remains to show that $\bar{S}$ converges to a KKT point of~\eqref{eqn:bmip} if the set of KKT points of~\eqref{eqn:bmip} is finite. This is in fact a straightforward corollary of~\cite[Theorem~4.3]{dinh2011combining}, by noticing that the assumptions thereof can be readily verified. For simplicity, we highlight the validity of only a few of these assumptions: Since $\zz^0$ in Algorithm~\refeq{alg:localBMI} is a strictly feasible solution to~\eqref{eqn:bmip}, the relative interior of the feasible set of~\eqref{eqn:bmip} is non-empty and thus Assumption~A1 in~\cite{dinh2011combining} holds; Our Assumption~\ref{ass:compactness} on the boundedness of the search space ensures that $g(\zz)$ in~\eqref{eqn:bmip} is bounded from above over a bounded feasible set, and therefore the boundedness assumptions in~\cite[Theorem~4.3]{dinh2011combining} holds.
	\qed
\end{proof}

\begin{proof}[of Lemma~\ref{lem:LMIsolution}]
	Let $\Lambda_\aaa \define \min_{1 \le \iota \le l} -\rho\left(\mathcal{F}_\iota(\aaa, \sss)\big\vert_{\sss = \left(c_\iota, 0, \ldots, 0\right)}\right)$, where $\rho(A)$ denotes the spectral radius of matrix $A$, i.e., the largest absolute value of the eigenvalues of $A$. It follows that program~\eqref{eqn:lmiExponetialBc} has a strictly feasible solution if $\lambda < \Lambda_\aaa$.
	
	Furthermore, under Assumption~\ref{ass:compactness} on the boundedness of parameter $\aaa \in \compactSet_{\aaa}$, $\Lambda_\aaa$ can be shown to be bounded by the well-known Gershgorin circle theorem. 
	
	Therefore, by taking an interior point of $\compactSet_{\aaa}$ as $\tilde{\aaa}$, and $\tilde{\lambda} = \Lambda_{\tilde{\aaa}} - \epsilon$ for some $\epsilon \in \RR^+$, we obtain a strictly feasible solution $(\tilde{\lambda}, \tilde{\aaa})$ to program~\eqref{eqn:lmiExponetialBc}. 
	\qed
\end{proof}

\begin{proof}[of Theorem~\ref{thm:bbCompleteness}]
    When the assumptions (1)--(3) hold, Algorithm~\ref{alg:bbAlgorithm} will eventually visit a branch wherein any parameter is valid (in case a valid parameter has not been found yet). If the necessary condition for invariant barrier certificates in Theorem~\ref{thm:invariantBcSosNecessary} is used to form the BMI problem~\eqref{eqn:bmiBc}, Line~\ref{lin:checkForNec} ensures to return a valid parameter $\bar{\aaa} \in \compactSet_{\aaa}$; Otherwise if the BMI problem~\eqref{eqn:bmiBc} encodes the sufficient condition in Theorem~\ref{thm:invariantBcSosSufficient} which strengthens the invariant barrier-certificate condition in Definition~\ref{def:invBc}, a valid parameter $\bar{\aaa}$ may not induce a non-negative objective value of~\eqref{eqn:bmiBc}. In this case, however, any parameter sampled and returned by Line~\ref{lin:startSample}--\ref{lin:endSample} in the branch is valid, as it contains only valid parameters.
	\qed
\end{proof}

\section{Benchmark Examples}\label{appendix_examples}

\begin{figure}[htbp]
	\centering
	\resizebox{\textwidth}{!}{
		\begin{tabular}{ccc}
			\subfloat[\expname{barr-cert1}]{~~~~\includegraphics[scale=0.28]{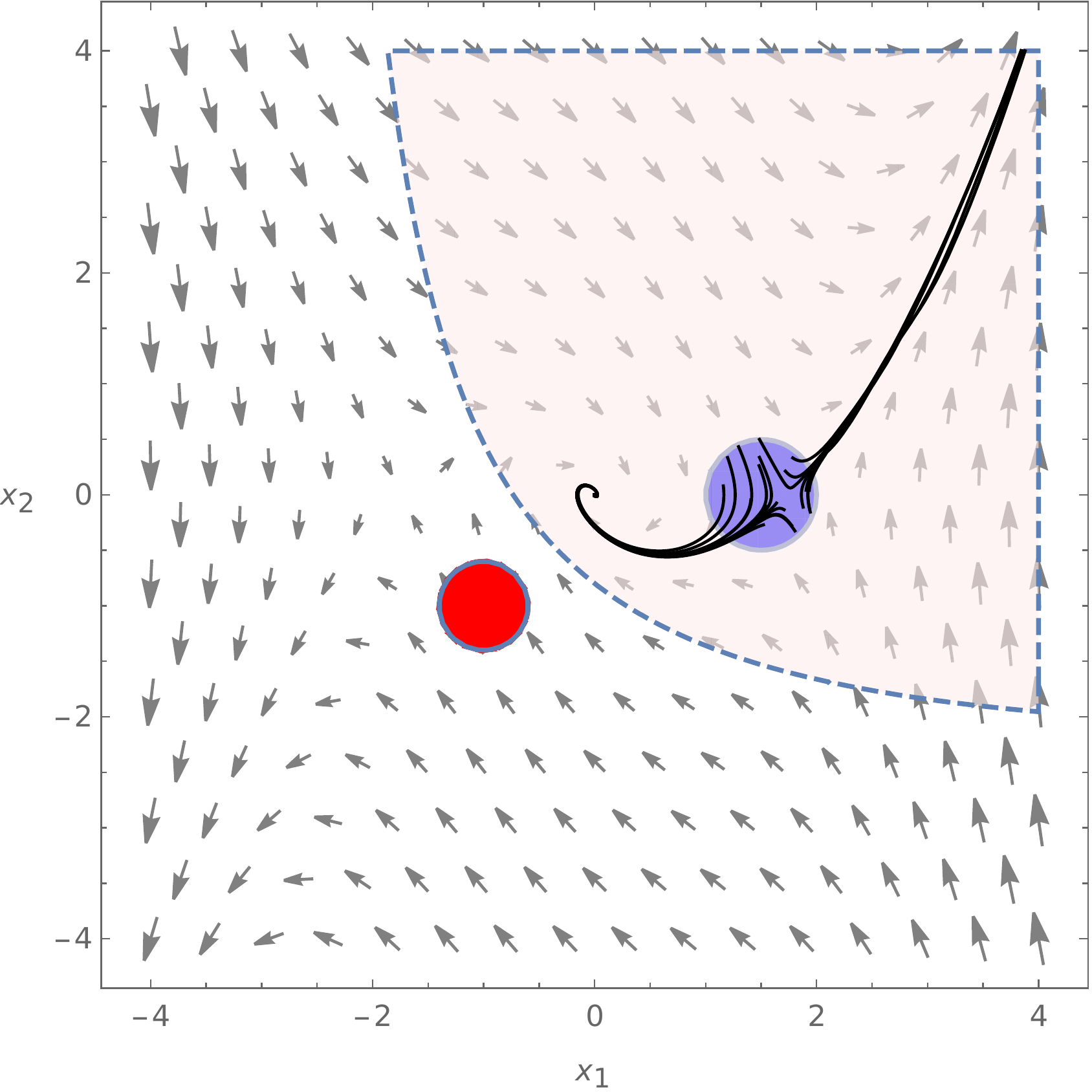}~~~~\label{fig:exmp1}}& 
			\subfloat[\expname{lie-der}]{~~~~\includegraphics[scale=0.28]{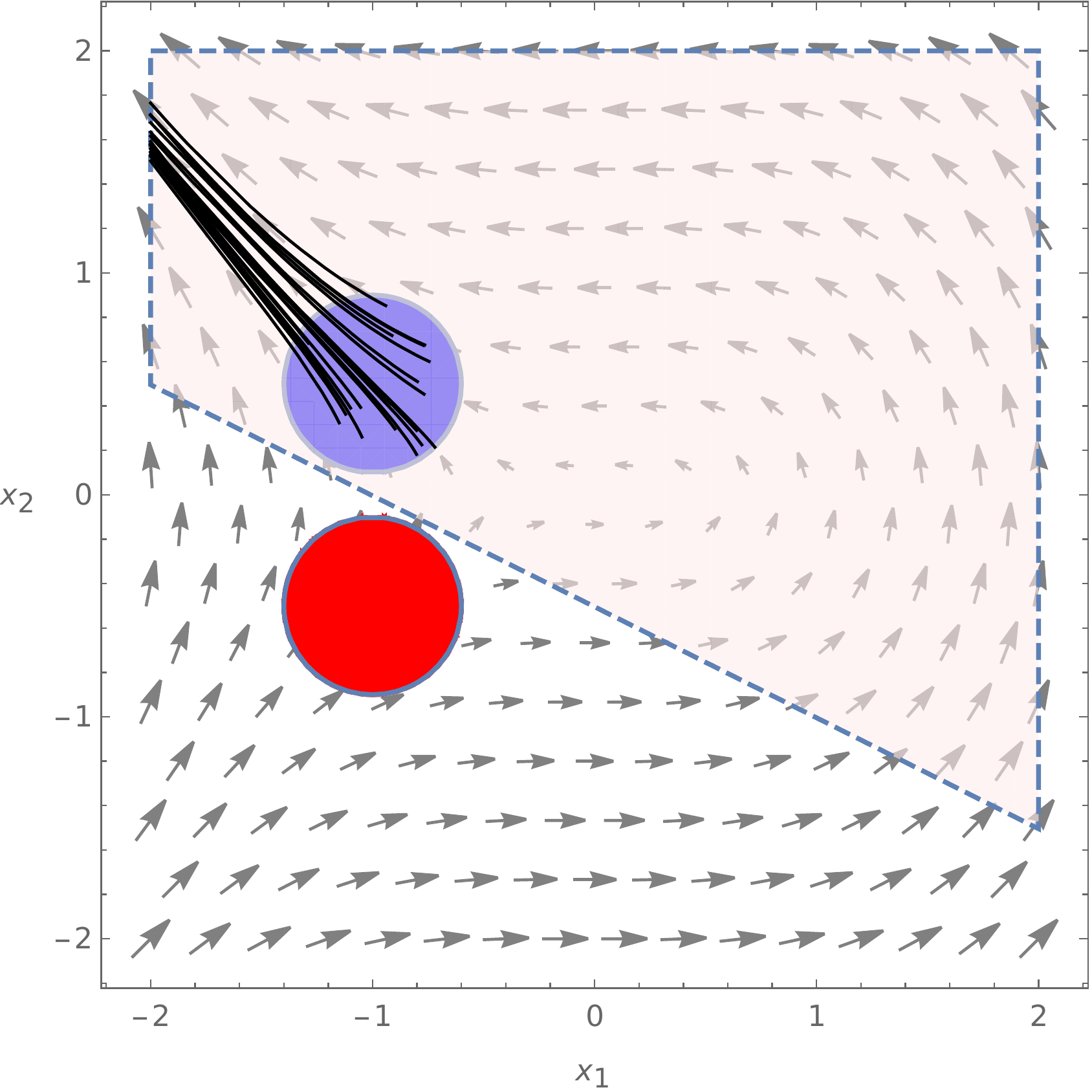}~~~~\label{fig:exmp8}}& 
			\hspace*{.1cm}
			\subfloat[\expname{stabilization}]{~~~~\includegraphics[scale=0.382]{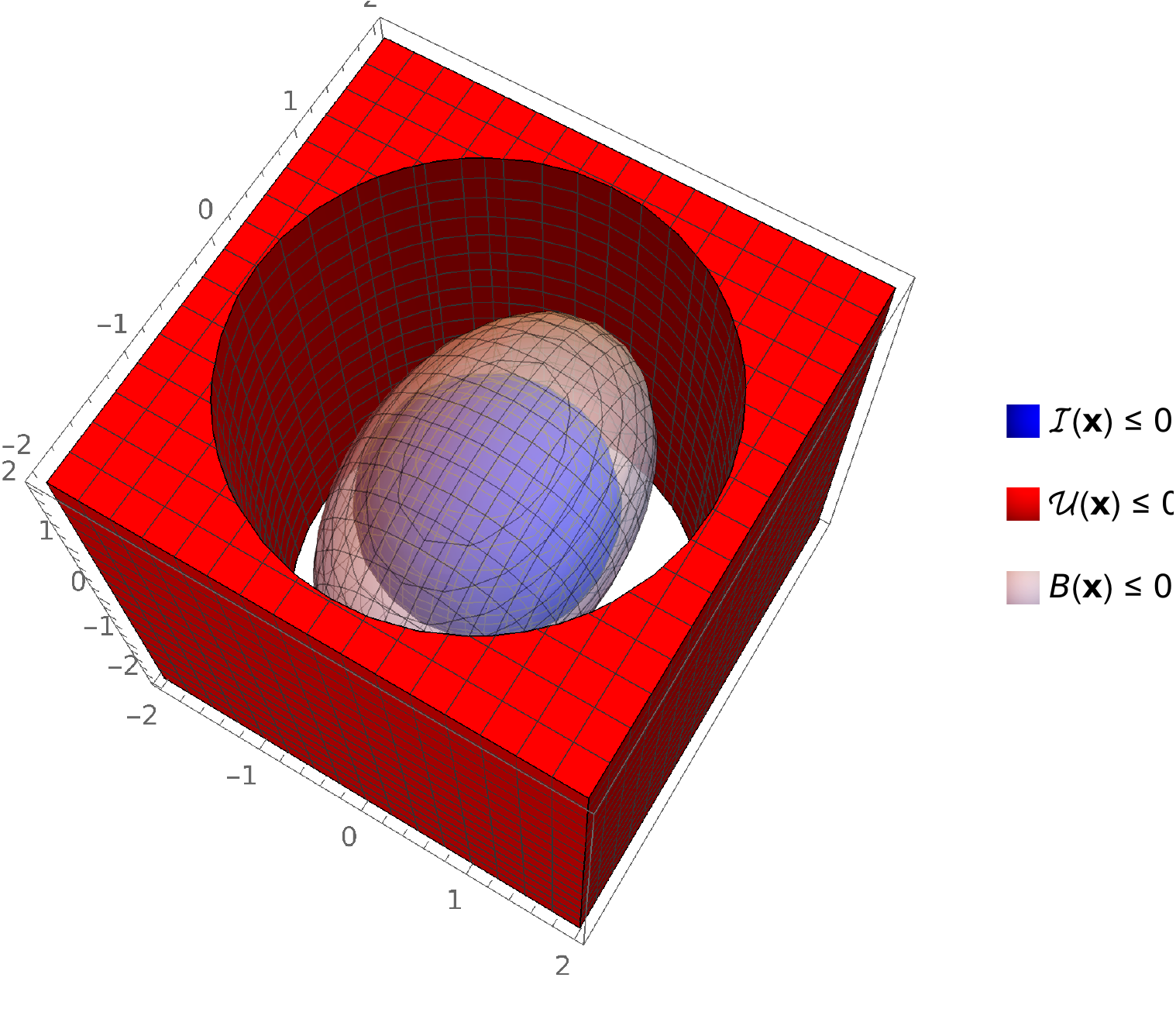}~~~~\label{fig:C14}}
			\hspace*{-.1cm}
			\\[-.1cm]
			
			\subfloat[\expname{clock}]{~~~~\includegraphics[scale=0.28]{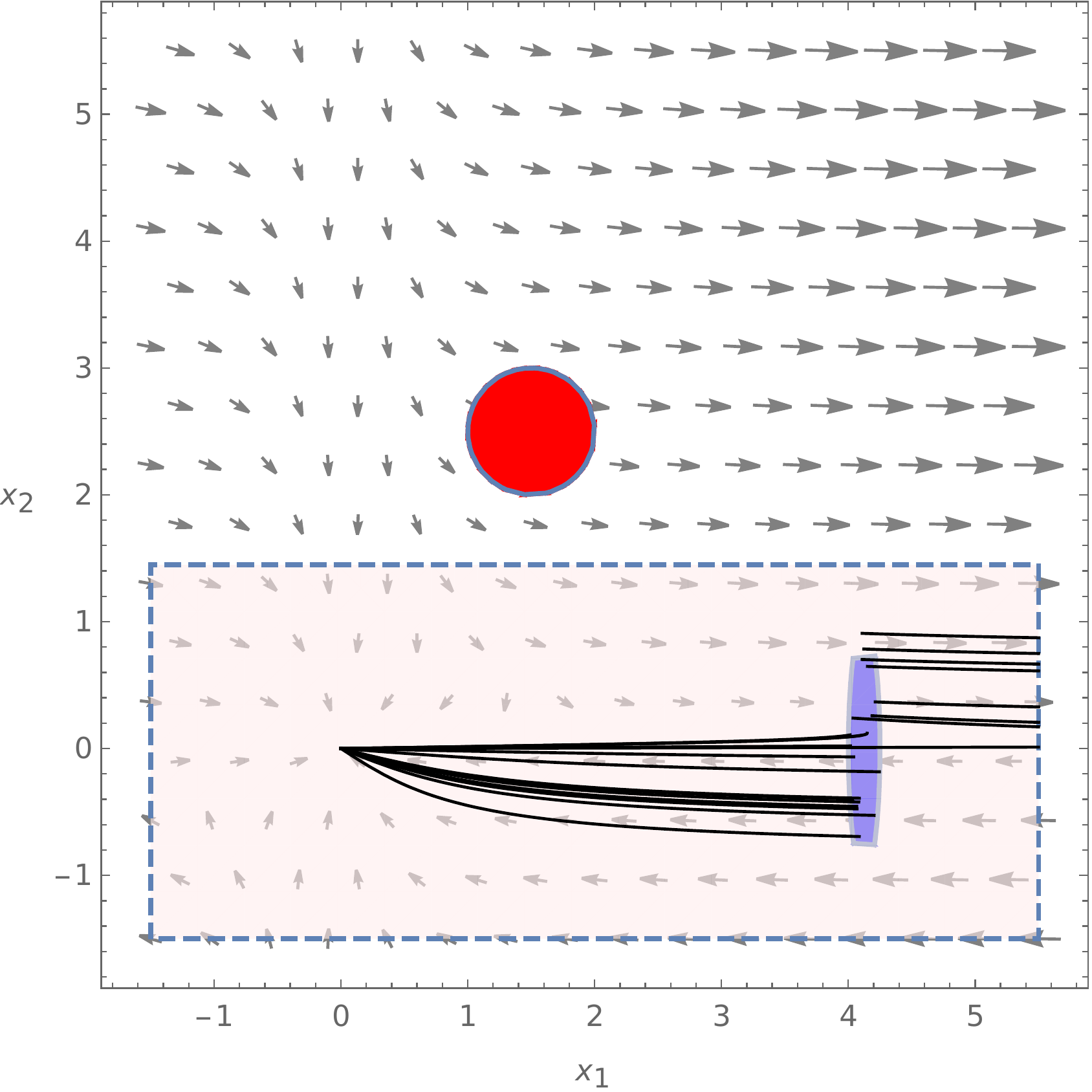}~~~~\label{fig:C1}}& 
			\subfloat[\expname{arch3}]{~~~~\includegraphics[scale=0.28]{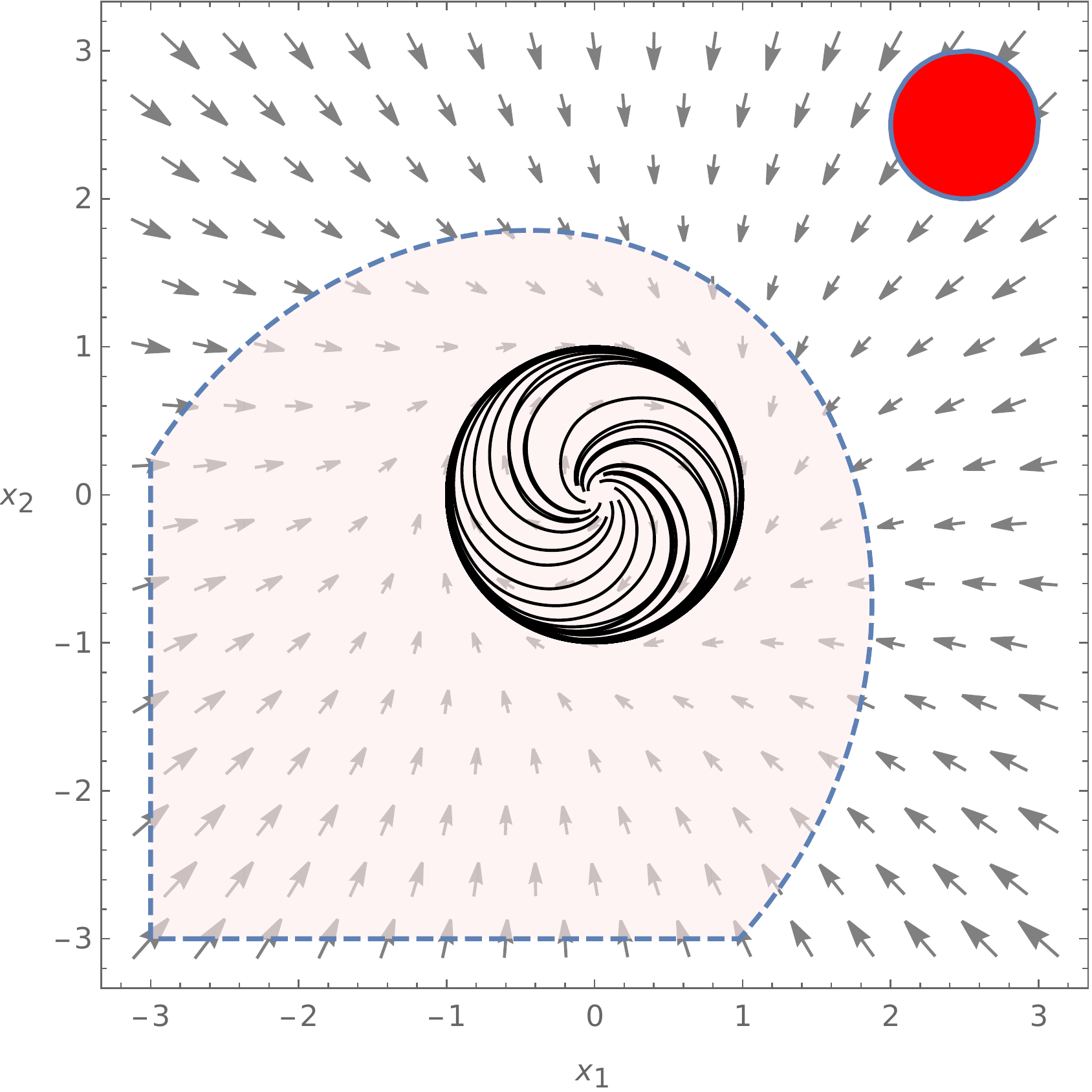}~~~~\label{fig:C5}}& 
			\subfloat[\expname{fitzhugh-nagumo}]{~~~~\includegraphics[scale=0.378]{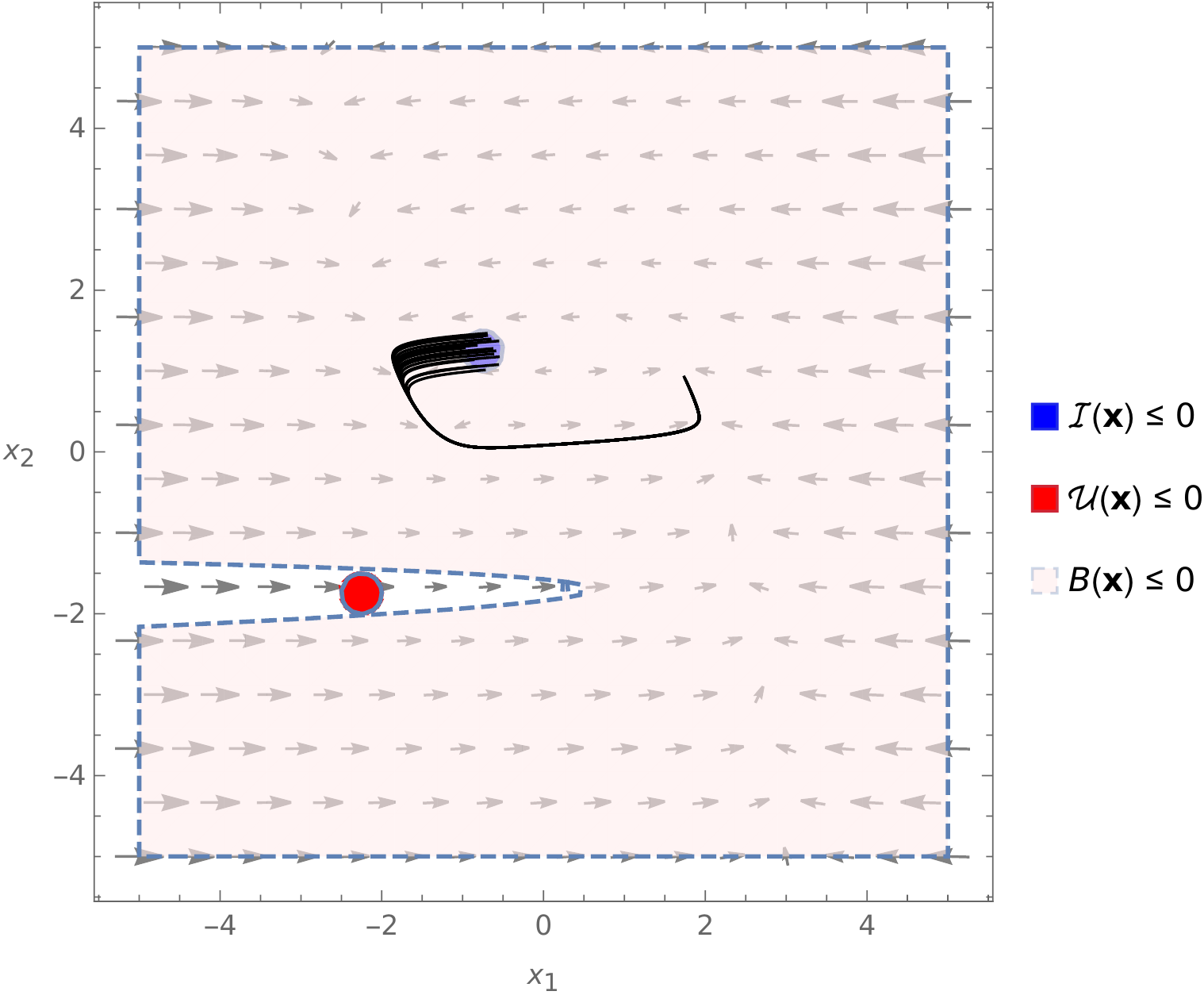}~~~~\label{fig:C12}}\\[-.1cm]
		\end{tabular}
	}
	\caption{Phase portraits of another selected set of examples with the synthesized invariant barrier certificates. The arrows indicate the vector field (hidden in 3D-graphics for a clear presentation) and the solid curves are randomly sampled trajectories.}\label{fig:visualization1}
\end{figure}

\begin{example}[\expname{contrived}]
    The vector flow field is: 
    \begin{equation*}
	\dot{\xx} =
    \begin{pmatrix}
        \dot{x}_1 \\
        \dot{x}_2
    \end{pmatrix} 
    =
    \begin{pmatrix}
        -x_1 + x_2 \\
        -x_2 
    \end{pmatrix}
	.
\end{equation*}

\begin{itemize}
    \item $\init = \{ \xx \in \mathbb{R}^2 \mid (x_1 - 1.125)^2 + (x_2 - 0.625)^2 - 0.0125 \leq 0 \}$. 
    \item $\unsafe = \{ \xx \in \mathbb{R}^2 \mid (x_1 - 0.875)^2 + (x_2 - 0.125)^2 - 0.0125 \leq 0 \}$. 
    \item $\domain = \{ \xx \in \mathbb{R}^2 \mid 0 \leq x_1, x_2 \leq 2 \}$. 
    \item $B(\aaa, \xx)$ includes all monomials up to degree $2$. 
\end{itemize}
\end{example}

\begin{example}[\expname{lie-der}~\textnormal{\cite{LZZ11}}]
    The vector flow field is: 
    \begin{equation*}
	\dot{\xx} =
    \begin{pmatrix}
        \dot{x}_1 \\
        \dot{x}_2
    \end{pmatrix} 
    =
    \begin{pmatrix}
        -2 x_2 \\
        x_1^2
    \end{pmatrix}
	.
\end{equation*}

\begin{itemize}
    \item $\init = \{ \xx \in \mathbb{R}^2 \mid (x_1 + 1)^2 + (x_2 - 0.5)^2 - 0.16 \leq 0 \}$. 
    \item $\unsafe = \{ \xx \in \mathbb{R}^2 \mid (x_1 + 1)^2 + (x_2 + 0.5)^2 - 0.16 \leq 0 \}$. 
    \item $\domain = \{ \xx \in \mathbb{R}^2 \mid -2 \leq x_1, x_2 \leq 2 \}$. 
    \item $B(\aaa, \xx)$ includes all monomials up to degree $1$. 
\end{itemize}
\end{example}

\begin{example}[\expname{lorenz}~\textnormal{\cite{djaballah2017construction}}]
    The vector flow field is: 
    \begin{equation*}
	\dot{\xx} =
    \begin{pmatrix}
        \dot{x}_1 \\
        \dot{x}_2 \\
        \dot{x}_3
    \end{pmatrix} 
    =
    \begin{pmatrix}
        10.0 (-x_1 + x_2) \\
        -x_2 + x_1 (28.0 - x_3) \\
        x_1 x_2 - \frac{8}{3} x_3
    \end{pmatrix}
	.
\end{equation*}

\begin{itemize}
    \item $\init = \{ \xx \in \mathbb{R}^3 \mid (x_1 + 14.5)^2 + (x_2 + 14.5)^2 + (x_3 - 12.5)^2 - 0.25 \leq 0 \}$. 
    \item $\unsafe = \{ \xx \in \mathbb{R}^3 \mid (x_1 + 16.5)^2 + (x_2 + 14.5)^2 + (x_3 - 2.5)^2 - 0.25 \leq 0 \}$. 
    \item $\domain = \{ \xx \in \mathbb{R}^3 \mid -20 \leq x_1, x_2, x_3 \leq 20 \}$. 
    \item $B(\aaa, \xx)$ includes all monomials up to degree $2$. 
\end{itemize}
\end{example}

\begin{example}[\expname{lti-stable}~\textnormal{\cite{DBLP:conf/cav/GaoKDRSAK19}}]
    The vector flow field is: 
    \begin{equation*}
	\dot{\xx} =
    \begin{pmatrix}
        \dot{x}_1 \\
        \dot{x}_2
    \end{pmatrix} 
    =
    \begin{pmatrix}
        -0.1 x_1 - 10 x_2 \\
        4 x_1 - 2 x_2 
    \end{pmatrix}
	.
\end{equation*}

\begin{itemize}
    \item $\init = \{ \xx \in \mathbb{R}^2 \mid (x_1 - 1.125)^2 + (x_2 - 0.625)^2 - 0.125^2 \leq 0 \}$. 
    \item $\unsafe = \{ \xx \in \mathbb{R}^2 \mid (x_1 + 1.5)^2 + (x_2 + 1.25)^2 - 0.25^2 \leq 0 \}$. 
    \item $\domain = \{ \xx \in \mathbb{R}^2 \mid -2 \leq x_1, x_2 \leq 2 \}$. 
    \item $B(\aaa, \xx)$ includes all monomials up to degree $2$. 
\end{itemize}
\end{example}

\begin{example}[\expname{lotka-volterra}~\textnormal{\cite{goubault2014finding}}]
    The vector flow field is: 
    \begin{equation*}
	\dot{\xx} =
    \begin{pmatrix}
        \dot{x}_1 \\
        \dot{x}_2 \\
        \dot{x}_3
    \end{pmatrix} 
    =
    \begin{pmatrix}
        x_1 (1 - x_3) \\
        x_2 (1 - 2 x_3) \\
        x_3 (-1 + x_1 + x_2)
    \end{pmatrix}
	.
\end{equation*}

\begin{itemize}
    \item $\init = \{ \xx \in \mathbb{R}^3 \mid (x_1 - 1)^2 + (x_2 - 1)^2 + x_3^2 - 0.64 \leq 0 \}$. 
    \item $\unsafe = \{ \xx \in \mathbb{R}^3 \mid x_1^2 + (x_2 + 1)^2 - 0.25 \leq 0 \}$. 
    \item $\domain = \{ \xx \in \mathbb{R}^3 \mid -2 \leq x_1, x_2, x_3 \leq 2 \}$. 
    \item $B(\aaa, \xx) = a x_2$. 
\end{itemize}
\end{example}

\begin{example}[\expname{clock}~\textnormal{\cite{RatschanS05}}]
    The vector flow field is: 
    \begin{equation*}
	\dot{\xx} =
    \begin{pmatrix}
        \dot{x}_1 \\
        \dot{x}_2
    \end{pmatrix} 
    =
    \begin{pmatrix}
        -x_1 + 2 x_1^2 x_2 \\
        -x_2 
    \end{pmatrix}
	.
\end{equation*}

\begin{itemize}
    \item $\init = \{ \xx \in \mathbb{R}^2 \mid (8 x_1 - 33)^2 + x_2^2 - 1 \leq 0 \}$. 
    \item $\unsafe = \{ \xx \in \mathbb{R}^2 \mid (x_1 - 1.5)^2 + (x_2 - 2.5)^2 - 0.25 \leq 0 \}$. 
    \item $\domain = \{ \xx \in \mathbb{R}^2 \mid -1.5 \leq x_1, x_2 \leq 5.5 \}$. 
    \item $B(\aaa, \xx)$ includes all monomials up to degree $1$. 
\end{itemize}
\end{example}

\begin{example}[\expname{lyapunov}~\textnormal{\cite{ratschan2010providing}}]
    The vector flow field is: 
    \begin{equation*}
	\dot{\xx} =
    \begin{pmatrix}
        \dot{x}_1 \\
        \dot{x}_2 \\
        \dot{x}_3
    \end{pmatrix} 
    =
    \begin{pmatrix}
        -x_2 \\
        -x_3 \\
        -x_1 - 2 x_2 - x_3 + x_1^3 
    \end{pmatrix}
	.
\end{equation*}

\begin{itemize}
    \item $\init = \{ \xx \in \mathbb{R}^3 \mid (x_1 - 0.25)^2 + (x_2 - 0.25)^2 + (x_3 - 0.25)^2 - 0.25 \leq 0 \}$. 
    \item $\unsafe = \{ \xx \in \mathbb{R}^3 \mid (x_1 - 1.5)^2 + (x_2 + 1.5)^2 + (x_3 + 1.5)^2 - 0.25 \leq 0 \}$. 
    \item $\domain = \{ \xx \in \mathbb{R}^3 \mid -2 \leq x_1, x_2, x_3 \leq 2 \}$. 
    \item $B(\aaa, \xx)$ includes all monomials up to degree $2$. 
\end{itemize}
\end{example}

\begin{example}[\expname{arch1}~\textnormal{\cite{sogokon2016non}}]
    The vector flow field is: 
    \begin{equation*}
	\dot{\xx} =
    \begin{pmatrix}
        \dot{x}_1 \\
        \dot{x}_2
    \end{pmatrix} 
    =
    \begin{pmatrix}
        -x_1 + 2 x_1^3 x_2^2 \\
        -x_2 
    \end{pmatrix}
	.
\end{equation*}

\begin{itemize}
    \item $\init = \{ \xx \in \mathbb{R}^2 \mid x_1^2 + (x_2 - 0.5)^2 - 0.04 \leq 0 \}$. 
    \item $\unsafe = \{ \xx \in \mathbb{R}^2 \mid (x_1 + 1.5)^2 + (x_2 + 1.5)^2 - 0.25 \leq 0 \}$. 
    \item $\domain = \{ \xx \in \mathbb{R}^2 \mid -2 \leq x_1, x_2 \leq 2 \}$. 
    \item $B(\aaa, \xx)$ includes all monomials up to degree $2$. 
\end{itemize}
\end{example}

\begin{example}[\expname{arch2}~\textnormal{\cite{sogokon2016non}}]
    The vector flow field is: 
    \begin{equation*}
	\dot{\xx} =
    \begin{pmatrix}
        \dot{x}_1 \\
        \dot{x}_2
    \end{pmatrix} 
    =
    \begin{pmatrix}
        x_1^2 + x_2^2 - 1\\
        5(x_1 x_2 - 1)
    \end{pmatrix}
	.
\end{equation*}

\begin{itemize}
    \item $\init = \{ \xx \in \mathbb{R}^2 \mid (x_1 + 0.5)^2 + (x_2 + 0.5)^2 - 0.25 \leq 0 \}$. 
    \item $\unsafe = \{ \xx \in \mathbb{R}^2 \mid (x_1 + 1.5)^2 + (x_2 + 1.5)^2 - 0.25 \leq 0 \}$. 
    \item $\domain = \{ \xx \in \mathbb{R}^2 \mid -2 \leq x_1, x_2 \leq 2 \}$. 
    \item $B(\aaa, \xx)$ includes all monomials up to degree $2$. 
\end{itemize}
\end{example}

\begin{example}[\expname{arch3}~\textnormal{\cite{sogokon2016non}}]
    The vector flow field is: 
    \begin{equation*}
	\dot{\xx} =
    \begin{pmatrix}
        \dot{x}_1 \\
        \dot{x}_2
    \end{pmatrix} 
    =
    \begin{pmatrix}
        x_1 - x_1^3 + x_2 - x_1 x_2^2 \\
        -x_1 + x_2 - x_1^2 x_2 - x_2^3
    \end{pmatrix}
	.
\end{equation*}

\begin{itemize}
    \item $\init = \{ \xx \in \mathbb{R}^2 \mid x_1^2 + x_2^2 - 0.04 \leq 0 \}$. 
    \item $\unsafe = \{ \xx \in \mathbb{R}^2 \mid (x_1 - 2.5)^2 + (x_2 - 2.5)^2 - 0.25 \leq 0 \}$. 
    \item $\domain = \{ \xx \in \mathbb{R}^2 \mid -3 \leq x_1, x_2 \leq 3 \}$. 
    \item $B(\aaa, \xx)$ includes all monomials up to degree $2$. 
\end{itemize}
\end{example}

\begin{example}[\expname{arch4}~\textnormal{\cite{sogokon2016non}}]
    The vector flow field is: 
    \begin{equation*}
	\dot{\xx} =
    \begin{pmatrix}
        \dot{x}_1 \\
        \dot{x}_2
    \end{pmatrix} 
    =
    \begin{pmatrix}
        -2 x_1 + x_1^2 + x_2 \\
        x_1 - 2 x_2 + x_2^2
    \end{pmatrix}
	.
\end{equation*}

\begin{itemize}
    \item $\init = \{ \xx \in \mathbb{R}^2 \mid x_1^2 + x_2^2 - 0.1^2 \leq 0 \}$. 
    \item $\unsafe = \{ \xx \in \mathbb{R}^2 \mid (x_1 - 0.75)^2 + (x_2 - 0.75)^2 - 0.25^2 \leq 0 \}$. 
    \item $\domain = \{ \xx \in \mathbb{R}^2 \mid -0.5 \leq x_1, x_2 \leq 1 \}$. 
    \item $B(\aaa, \xx)$ includes all monomials up to degree $1$. 
\end{itemize}
\end{example}

\begin{example}[\expname{barr-cert1}~\textnormal{\cite{Prajna04}}]
    The vector flow field is: 
    \begin{equation*}
	\dot{\xx} =
    \begin{pmatrix}
        \dot{x}_1 \\
        \dot{x}_2
    \end{pmatrix} 
    =
    \begin{pmatrix}
        x_2 \\
        -x_1 + \frac{1}{3} x_1^3 - x_2 
    \end{pmatrix}
	.
\end{equation*}

\begin{itemize}
    \item $\init = \{ \xx \in \mathbb{R}^2 \mid (x_1 - 1.5)^2 + x_2^2 - 0.25 \leq 0 \}$. 
    \item $\unsafe = \{ \xx \in \mathbb{R}^2 \mid (x_1 + 1)^2 + (x_2 + 1)^2 - 0.16 \leq 0 \}$. 
    \item $\domain = \{ \xx \in \mathbb{R}^2 \mid -4 \leq x_1, x_2 \leq 4 \}$. 
    \item $B(\aaa, \xx)$ includes all monomials up to degree $2$. 
\end{itemize}
\end{example}

\begin{example}[\expname{barr-cert2}~\textnormal{\cite{djaballah2017construction}}]
    The vector flow field is: 
    \begin{equation*}
	\dot{\xx} =
    \begin{pmatrix}
        \dot{x}_1 \\
        \dot{x}_2
    \end{pmatrix} 
    =
    \begin{pmatrix}
        -x_1 + x_1 x_2 \\
        -x_2 
    \end{pmatrix}
	.
\end{equation*}

\begin{itemize}
    \item $\init = \{ \xx \in \mathbb{R}^2 \mid (x_1 - 1.125)^2 + (x_2 - 0.625)^2 - 0.125^2 \leq 0 \}$. 
    \item $\unsafe = \{ \xx \in \mathbb{R}^2 \mid (x_1 - 0.875)^2 + (x_2 - 0.125)^2 - 0.075^2 \leq 0 \}$. 
    \item $\domain = \{ \xx \in \mathbb{R}^2 \mid 0 \leq x_1, x_2 \leq 1.5 \}$. 
    \item $B(\aaa, \xx)$ includes all monomials up to degree $2$. 
\end{itemize}
\end{example}

\begin{example}[\expname{barr-cert3}~\textnormal{\cite{zhang2018safety}}]
    The vector flow field is: 
    \begin{equation*}
	\dot{\xx} =
    \begin{pmatrix}
        \dot{x}_1 \\
        \dot{x}_2
    \end{pmatrix} 
    =
    \begin{pmatrix}
        -x_1 + x_1 x_2 \\
        -x_2 
    \end{pmatrix}
	.
\end{equation*}

\begin{itemize}
    \item $\init = \{ \xx \in \mathbb{R}^2 \mid (x_1 + 1)^2 + (x_2 + 1)^2 - 0.25 \leq 0 \}$. 
    \item $\unsafe = \{ \xx \in \mathbb{R}^2 \mid x_1^2 + (x_2 - 1)^2 - 0.25 \leq 0 \}$. 
    \item $\domain = \{ \xx \in \mathbb{R}^2 \mid -2 \leq x_1, x_2 \leq 2 \}$. 
    \item $B(\aaa, \xx)$ includes all monomials up to degree $1$. 
\end{itemize}
\end{example}

\begin{example}[\expname{barr-cert4}~\textnormal{\cite{zhang2018safety}}]
    The vector flow field is: 
    \begin{equation*}
	\dot{\xx} =
    \begin{pmatrix}
        \dot{x}_1 \\
        \dot{x}_2
    \end{pmatrix} 
    =
    \begin{pmatrix}
        -x_1 + 2 x_1^2 x_2 \\
        -x_2 
    \end{pmatrix}
	.
\end{equation*}

\begin{itemize}
    \item $\init = \{ \xx \in \mathbb{R}^2 \mid 9 x_1^2 + (2 x_2 - 2.25)^2 - 0.75^2 \leq 0 \}$. 
    \item $\unsafe = \{ \xx \in \mathbb{R}^2 \mid (x_1 - 2)^2 + (x_2 - 2)^2 - 0.5^2 \leq 0 \}$. 
    \item $\domain = \{ \xx \in \mathbb{R}^2 \mid -1 \leq x_1, x_2 \leq 3 \}$. 
    \item $B(\aaa, \xx)$ includes all monomials up to degree $2$. 
\end{itemize}
\end{example}

\begin{example}[\expname{fitzhugh-nagumo}~\textnormal{\cite{DBLP:conf/cdc/SassiGS14}}]
    The vector flow field is: 
    \begin{equation*}
	\dot{\xx} =
    \begin{pmatrix}
        \dot{x}_1 \\
        \dot{x}_2
    \end{pmatrix} 
    =
    \begin{pmatrix}
        -1/3 x_1^3 + x_1 - x_2 + 0.875 \\
        0.08 (x_1 - 0.8 x_2 + 0.7)
    \end{pmatrix}
	.
\end{equation*}

\begin{itemize}
    \item $\init = \{ \xx \in \mathbb{R}^2 \mid (x_1 + 0.75)^2 + (x_2 -1.25)^2 - 0.25^2 \leq 0 \}$. 
    \item $\unsafe = \{ \xx \in \mathbb{R}^2 \mid (x_1 + 2.25)^2 + (x_2 + 1.75)^2 - 0.25^2 \leq 0 \}$. 
    \item $\domain = \{ \xx \in \mathbb{R}^2 \mid -5 \leq x_1, x_2 \leq 5 \}$. 
    \item $B(\aaa, \xx)$ includes all monomials up to degree $2$. 
\end{itemize}
\end{example}

\begin{example}[\expname{stabilization}~\textnormal{\cite{DBLP:conf/hybrid/SassiS15}}]
    The vector flow field is: 
    \begin{equation*}
	\dot{\xx} =
    \begin{pmatrix}
        \dot{x}_1 \\
        \dot{x}_2 \\
        \dot{x}_3
    \end{pmatrix} 
    =
    \begin{pmatrix}
        -x_1 + x_2 - x_3 \\
        -x_1 (x_3 + 1) - x_2 \\
        0.76524 x_1 - 4.7037 x_3
    \end{pmatrix}
	.
\end{equation*}

\begin{itemize}
    \item $\init = \{ \xx \in \mathbb{R}^3 \mid x_1^2 + x_2^2 + x_3^2 - 1 \leq 0 \}$. 
    \item $\unsafe = \{ \xx \in \mathbb{R}^3 \mid -x_1^2 - x_2^2 + 3 \leq 0 \}$. 
    \item $\domain = \{ \xx \in \mathbb{R}^3 \mid -2 \leq x_1, x_2, x_3 \leq 2 \}$. 
    \item $B(\aaa, \xx)$ includes all monomials up to degree $2$. 
\end{itemize}
\end{example}

\begin{example}[\expname{lie-high-order}]
	The vector flow field is: 
	\begin{equation*}
		\dot{\xx} =
		\begin{pmatrix}
			\dot{x}_1 \\
			\dot{x}_2 
		\end{pmatrix} 
		=
		\begin{pmatrix}
			x_1 \\
			x_2
		\end{pmatrix}
		.
	\end{equation*}
	
	\begin{itemize}
		\item $\init = \{ \xx \in \mathbb{R}^2 \mid (x_1-1.125)^2+(x_2-0.625)^2-0.0125 \leq 0 \}$. 
		\item $\unsafe = \{ \xx \in \mathbb{R}^2 \mid (x_1-0.875)^2+(x_2-0.125)^2-0.0125 \leq 0 \}$. 
		\item $\domain = \{ \xx \in \mathbb{R}^2 \mid -2 \leq x_1, x_2 \leq 2 \}$. 
		\item $B(\aaa, \xx) = x_1^2+ a_1 x_2^2 + a_2 x_1+ a_3 x_2+a_4$. 
	\end{itemize}
\end{example}

\begin{example}[\expname{raychaudhuri}~\textnormal{\cite{ferragut2015seeking}}]
    The vector flow field is: 
    \begin{equation*}
	\dot{\xx} =
    \begin{pmatrix}
        \dot{x}_1 \\
        \dot{x}_2 \\
        \dot{x}_3 \\
        \dot{x}_4
    \end{pmatrix} 
    =
    \begin{pmatrix}
        -0.5 x_1^2 - 2 (x_2^2 + x_3^2 - x_4^2) \\
        -x_1 x_2 - 1 \\
        -x_1 x_3 \\
        -x_1 x_4
    \end{pmatrix}
	.
\end{equation*}

\begin{itemize}
    \item $\init = \{ \xx \in \mathbb{R}^4 \mid x_1^2 + (x_2 + 1)^2 - 0.1 \leq 0 \}$. 
    \item $\unsafe = \{ \xx \in \mathbb{R}^4 \mid (x_1 + 1)^2 + x_2^2 - 0.1 \leq 0 \}$. 
    \item $\domain = \{ \xx \in \mathbb{R}^4 \mid -1.5 \leq x_1, \ldots, x_4 \leq 1.5 \}$. 
    \item $B(\aaa, \xx) = a_1 x_1^2 + a_2 x_1 x_2 + a_3 x_2^2 + a_4 x_1 + a_5 x_2 + a_6$. 
\end{itemize}
\end{example}

\begin{example}[\expname{focus}~\textnormal{\cite{ratschan2006constraints}}]
    The vector flow field is: 
    \begin{equation*}
	\dot{\xx} =
    \begin{pmatrix}
        \dot{x}_1 \\
        \dot{x}_2
    \end{pmatrix} 
    =
    \begin{pmatrix}
        x_1 - x_2 \\
        x_1 + x_2 
    \end{pmatrix}
	.
\end{equation*}

\begin{itemize}
    \item $\init = \{ \xx \in \mathbb{R}^2 \mid (x_1 - 2.75)^2 + (5 x_2 - 10)^2 - 0.25^2 \leq 0 \}$. 
    \item $\unsafe = \{ \xx \in \mathbb{R}^2 \mid x_1 - 2 \leq 0 \}$. 
    \item $\domain = \{ \xx \in \mathbb{R}^2 \mid 1.5 \leq x_1, x_2 \leq 3.5 \}$. 
    \item $B(\aaa, \xx)$ includes all monomials up to degree $4$. 
\end{itemize}
\end{example}

\begin{example}[\expname{sys-bio1}~\textnormal{\cite{klipp2008systems}}]
    The vector flow field is: 
    \begin{equation*}
	\dot{\xx} =
    \begin{pmatrix}
        \dot{x}_1 \\
        \dot{x}_2 \\
        \dot{x}_3 \\
        \dot{x}_4 \\
        \dot{x}_5 \\
        \dot{x}_6 \\
        \dot{x}_7 
    \end{pmatrix} 
    =
    \begin{pmatrix}
        -0.4 x_1 + 5 x_3 x_4 \\
        0.4 x_1 - x_2 \\
        x_2 - 5 x_3 x_4 \\
        5 x_5 x_6 - 5 x_3 x_4 \\
        -5 x_5 x_6 + 5 x_3 x_4 \\
        0.5 x_7 - 5 x_5 x_6 \\
        -0.5 x_7 + 5 x_5 x_6
    \end{pmatrix}
	.
\end{equation*}

\begin{itemize}
    \item $\init = \{ \xx \in \mathbb{R}^7 \mid \sum_{i=1}^{7}(x_i - 1)^2 - 0.01^2 \leq 0 \}$. 
    \item $\unsafe = \{ \xx \in \mathbb{R}^7 \mid \sum_{i=1}^{7}(x_i - 1.9)^2 - 0.1^2 \leq 0 \}$. 
    \item $\domain = \{ \xx \in \mathbb{R}^7 \mid -2 \leq x_1, \ldots, x_7 \leq 2 \}$. 
    \item $B(\aaa, \xx)$ includes all monomials up to degree $2$. 
\end{itemize}
\end{example}

\begin{example}[\expname{sys-bio2}~\textnormal{\cite{klipp2008systems}}]
    The vector flow field is: 
    \begin{equation*}
	\dot{\xx} =
    \begin{pmatrix}
        \dot{x}_1 \\
        \dot{x}_2 \\
        \dot{x}_3 \\
        \dot{x}_4 \\
        \dot{x}_5 \\
        \dot{x}_6 \\
        \dot{x}_7 \\
        \dot{x}_8 \\
        \dot{x}_9 
    \end{pmatrix} 
    =
    \begin{pmatrix}
        3 x_3 - x_1 x_6 \\
        x_4 - x_2 x_6 \\
        x_1 x_6 - 3 x_3 \\
        x_2 x_6 - x_4 \\
        3 x_3 + 5 x_1 - x_5 \\
        5 x_5 + 3 x_3 + x_4 - x_6 (x_1 + x_2 + 2 x_8 + 1) \\
        5 x_4 + x_2 - 0.5 x_7 \\
        5 x_7 - 2 x_6 x_8 + x_9 - 0.2 x_8 \\
        2 x_6 x_8 - x_9
    \end{pmatrix}
	.
\end{equation*}

\begin{itemize}
    \item $\init = \{ \xx \in \mathbb{R}^9 \mid \sum_{i=1}^{9}(x_i - 1)^2 - 0.01^2 \leq 0 \}$. 
    \item $\unsafe = \{ \xx \in \mathbb{R}^9 \mid \sum_{i=1}^{9}(x_i - 1.9)^2 - 0.1^2 \leq 0 \}$. 
    \item $\domain = \{ \xx \in \mathbb{R}^9 \mid -2 \leq x_1, \ldots, x_9 \leq 2 \}$. 
    \item $B(\aaa, \xx)$ includes all monomials up to degree $1$. 
\end{itemize}
\end{example}

\begin{example}[\expname{quadcopter}~\textnormal{\cite{DBLP:conf/cav/GaoKDRSAK19}}]
    The vector flow field is: 
    \begin{equation*}
	\dot{\xx} =
    \begin{pmatrix}
        \dot{x}_1 \\
        \dot{x}_2 \\
        \dot{x}_3 \\
        \dot{x}_4 \\
        \dot{x}_5 \\
        \dot{x}_6 \\
        \dot{x}_7 \\
        \dot{x}_8 \\
        \dot{x}_9 \\
        \dot{x}_{10} \\
        \dot{x}_{11} \\
        \dot{x}_{12} 
    \end{pmatrix} 
    =
    \begin{pmatrix}
        x_4 \\
        x_5 \\
        x_6 \\
        -7253.4927 x_1 + 1936.3639 x_{11} - 1338.7624 x_4 + 1333.3333 x_8 \\
        -1936.3639 x_{10} - 7253.4927 x_2 - 1338.7624 x_5 - 1333.3333 x_7 \\
        -769.2308 x_3 - 770.2301 x_6 \\
        x_{10} \\
        x_{11} \\
        x_{12} \\
        9.81 x_2 \\
        -9.81 x_1 \\
        -16.3541 x_{12} - 15.3846 x_9
    \end{pmatrix}
	.
\end{equation*}

\begin{itemize}
    \item $\init = \{ \xx \in \mathbb{R}^{12} \mid \sum_{i=1}^{12} x_i^2 - 0.01 \leq 0 \}$. 
    \item $
    \begin{aligned}[t]
    \unsafe = \{ \xx \in \mathbb{R}^{12} \ & \mid (2 x_1 - 0.5)^2 + (2 x_2 - 0.5)^2 + (2 x_3 - 0.5)^2 + (x_4 - 1)^2 \\
    &+ (x_5 - 1)^2 + (x_6 - 1)^2 + (x_7 - 1)^2 + (x_8 + 1)^2 + (x_9 - 1)^2 \\
    &+ (x_{10} - 1)^2 + (x_{11} + 1)^2 + (x_{12} - 1)^2 - 0.25 \leq 0 \}.
    \end{aligned} $
    \item $\domain = \{ \xx \in \mathbb{R}^{12} \mid -2 \leq x_1, \ldots, x_{12} \leq 2 \}$. 
    \item $B(\aaa, \xx)$ includes all monomials up to degree $1$. 
\end{itemize}
\end{example}

\end{subappendices}

\end{document}